\documentclass[preprint,amsmath,amssymb]{revtex4}

\usepackage[dvips]{graphicx}
\usepackage{graphicx}
\usepackage{amsmath}
\usepackage{amssymb}
\usepackage{amsfonts}
\usepackage{upgreek}
\usepackage{txfonts}
\usepackage{color}
\usepackage{latexsym}
\usepackage{amsmath}
\usepackage{bm}
\usepackage{yfonts}
\usepackage{comment}

\usepackage{hyperref}
\usepackage{xcolor}
\definecolor{dark-red}{rgb}{0.4,0.15,0.15}
\definecolor{dark-blue}{rgb}{0.15,0.15,0.4}
\definecolor{medium-blue}{rgb}{0,0,0.5}
\hypersetup{
    colorlinks, linkcolor={dark-blue},
    citecolor={dark-blue}, urlcolor={medium-blue}
}

\begin{document}
\title{Dynamic criticality far-from-equilibrium:
one-loop flow of Burgers-Kardar-Parisi-Zhang 
systems with broken Galilean invariance}

\author{Philipp Strack}
\email{pstrack@physics.harvard.edu}
\homepage{http://users.physics.harvard.edu/~pstrack}

\affiliation{Institut f\"ur Theoretische Physik, Universit\"at zu K\"oln, D-50937 Cologne, Germany}
\affiliation{Department of Physics, Harvard University, Cambridge MA 02138}

\date{\today}

\begin{abstract}

Burgers-Kardar-Parisi-Zhang (KPZ) scaling has recently (re-) surfaced in a variety of 
physical contexts, ranging from anharmonic chains to quantum systems such as open 
superfluids, in which a variety of random forces may be encountered and/or engineered.
Motivated by these developments, we here provide a generalization of the KPZ universality 
class to situations with long-ranged temporal correlations in the noise, which purposefully 
break the Galilean invariance that is central to the conventional KPZ solution.
We compute the phase diagram and critical exponents of the KPZ equation with $1/f$-noise 
(KPZ$_{1/f}$) in spatial dimensions $1\leq d < 4$ using the dynamic renormalization group 
with a frequency cutoff technique in a one-loop truncation. Distinct features of KPZ$_{1/f}$ 
are: (i) a generically scale-invariant, rough phase at high noise levels that violates 
fluctuation-dissipation relations and exhibits hyperthermal statistics {\it even in d=1}, (ii) 
a fine-tuned roughening transition at which the flow fulfills an emergent thermal-like 
fluctuation-dissipation relation, that separates the rough phase from 
(iii) a {\it massive phase} in $1< d < 4$ (in $d=1$ the interface is always rough). 
We point out potential connections to nonlinear hydrodynamics with a reduced set of 
conservation laws and noisy quantum liquids.

\end{abstract}

\maketitle

-----------------------------------------------------------------------------------------------------------------\\[-15mm]
\tableofcontents

\newpage
\section{Introduction}

A lot of effort across the physical sciences is currently being 
directed at the derivation of the laws of statistical mechanics 
far-from-(thermal)-equilibrium \cite{jaeger10}. A prototypical question of interest 
begins with a many-particle system in a known initial state, whose 
statistics for example in energy space is known. Subsequently, 
the system is subjected to either a rapid change of its parameters 
or a non-equilibrium drive and/or dissipation. One then would like to 
understand how the statistical distributions evolve in time, in particular with regard to 
thermalization properties, that is, how quickly and by which mechanisms, 
energy/momenta are redistributed in phase space and real space. The steady-state 
distributions in the long-time limit $t\rightarrow \infty$ are also interesting.

Roughly speaking, there are two extreme scenarios: (i) Integrable systems with a 
large number of conservation laws whose thermalization is at least slow due to constraints 
in phase space from these conservation laws (see Ref.~\onlinecite{sotiriadis10} and 
references therein). (ii) Granular or soft matter systems (such as hard spheres in a box),
engineered liquids \cite{takeuchi10,yunker13}, or randomly sputtered interfaces \cite{kpz86}, 
in which the concept of temperature, a priori, 
does not make sense, there are typically fewer conserved quantities, and the dynamics is 
determined by geometric constraints, dimensionality, and/or the amount of disorder. Moreover, 
the amount of symmetry shapes the phase structure and associated transitions despite the 
absence of a well-defined free energy landscape far-from-equilibrium. 

In order to broadly elucidate the role of conservation laws in prototypical far-from-equilibrium 
phase transitions, the present paper studies 
Burgers-Kardar-Parisi-Zhang systems after {\it explicitly breaking its key symmetry/conservation law: 
the Galilean invariance} of the randomly stirred Burgers fluid associated with seeing the same 
physics when looking at the fluid in a moving frame \cite{forster77}.  
Another motivation is the possibility to learn about thermalization properties 
of quantum systems by finding ways to deform the 
``KPZ-attractive Lieb-Liniger bosons duality" \cite{kardar_bethe87,brunet00,calabrese11,calabrese14}
away from the integrable point/with less conservation laws.

In this paper, we perform a dynamic renormalization group analysis of the 
Burgers-Kardar-Parisi-Zhang equation subject to $1/f$-noise developing a frequency rescaling 
technique on the Keldysh contour.

\subsection{Model: Cole-Hopf transformed KPZ$_{1/f}$  equation}
\label{subsec:model}

Our calculations are based on the Cole-Hopf transformed KPZ equation (recapitulated below), 
when it becomes a gapless diffusion equation with multiplicative noise
\begin{align}
\gamma \partial_t \phi = \nu_0 \nabla^2 \phi + \frac{\lambda}{2\nu_0} \phi \eta\;
\label{eq:diffusion}
\end{align}
with $\phi = \phi(t,\mathbf{x})$ a scalar field describing fluctuations in time and space 
around a growing average height level, $\gamma$ a friction parameter, and $\nu_0$ 
the viscosity in the Burgers fluid picture. The $\eta$-field acts as nonlinear, multiplicative 
noise \cite{grinstein96} with coupling strength proportional to $\lambda$ \cite{medina89}. 
A simple way to break the Galilean invariance is to endow the noise with $1/f$-correlations 
in time such that in frequency representation
\begin{align}
\overline{\eta(\omega',\mathbf{x}')\eta(\omega,\mathbf{x})} 
=
D_{1/f}(\omega) \delta(\omega+\omega')
\delta^{(d)}(\mathbf{x}'-\mathbf{x})\;.
\label{eq:noise}
\end{align}
with ubiquitous $1/f$ or pink noise temporal correlations \cite{bak87}
\begin{align}
D_{1/f}(\omega)= \frac{1}{|\omega|}\;.
\label{eq:pink}
\end{align}
Several authors have considered temporally correlated noise in the 
context of KPZ \cite{medina89,lam92,halpin-healy95,katzav04} and, 
more recently, the $O(N)$-model \cite{bonart12}.
In particular, it was found that the conventional KPZ exponents 
may change as a consequence of correlations in the noise. 
However, within the dynamic renormalization group approach 
of Medina et al. \cite{medina89}, the interesting $1/f$-case 
could not be addressed due to the presence of infrared singularities 
also in the frequency integrations. It is one objective of the present 
paper to fill this gap.

Medina et al. \cite{medina89} have mentioned impurities 
or charged ions at the interface as possibility to generate 
temporal long-ranged correlations in the noise.
In the context of quantum systems, $1/f$ appears 
generically as charge noise in trapped ions for example
\cite{dalla_wrong,dalla12} and other types of noise can appear 
in the laser trapping potentials of ultra cold atoms \cite{buchhold14}.

\subsection{Key results}
\label{subsec:results}
\begin{figure} [t]
\includegraphics[width=85mm]{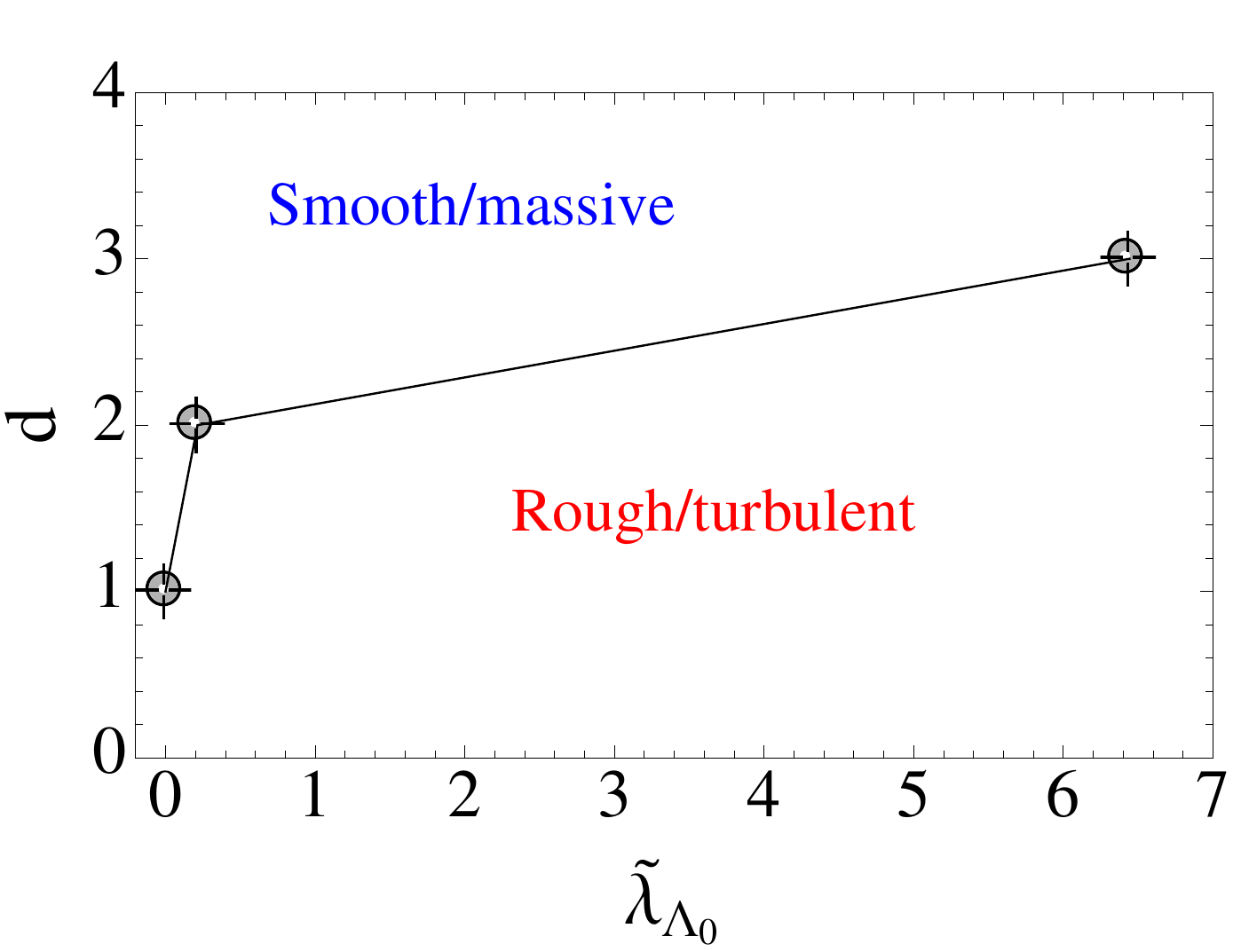}
\caption{Phase diagram of the KPZ equation with $1/f$ noise (computed points are connected 
as a guide to the eye) as a function of space dimensions ($d$) and strength of the bare noise vertex 
$\tilde{\lambda}_{\Lambda_0}$. The locators illustrate 
the roughening transition characterized by an unstable fixed point whose 
exponents fulfill an emergent thermal fluctuation-dissipation relation. In one dimension, 
the interface is always rough and any initial value of the noise vertex leads to 
the hyperthermal fixed point in the rough/turbulent phase. In contrast to conventional white noise KPZ 
phase diagram (e.g.: Ref.~\onlinecite{natter92}), 
here the fluctuation-dissipation theorem breaks down even in $d=1$ and the smooth phase is 
manifestly massive. 
}
\label{fig:phasediag}
\end{figure}
%


From the solution of our renormalization group equations, we obtain Fig.~\ref{fig:phasediag} as the phase 
diagram from a one-loop flow of Eq.~(\ref{eq:diffusion}) with $1/f$-noise integrating fluctuations from large 
frequencies $\omega=\Lambda_0$ to the lowest frequencies $\omega=0$.
In spatial dimensionality $d=2$, $d=3$, 
a massive phase for small $\tilde{\lambda}_{\Lambda_0}$
transits into a rough or turbulent phase at a critical value 
$\tilde{\lambda}_{\Lambda_0,c}$. 


%
\begin{figure} [t]
\includegraphics[width=80mm]{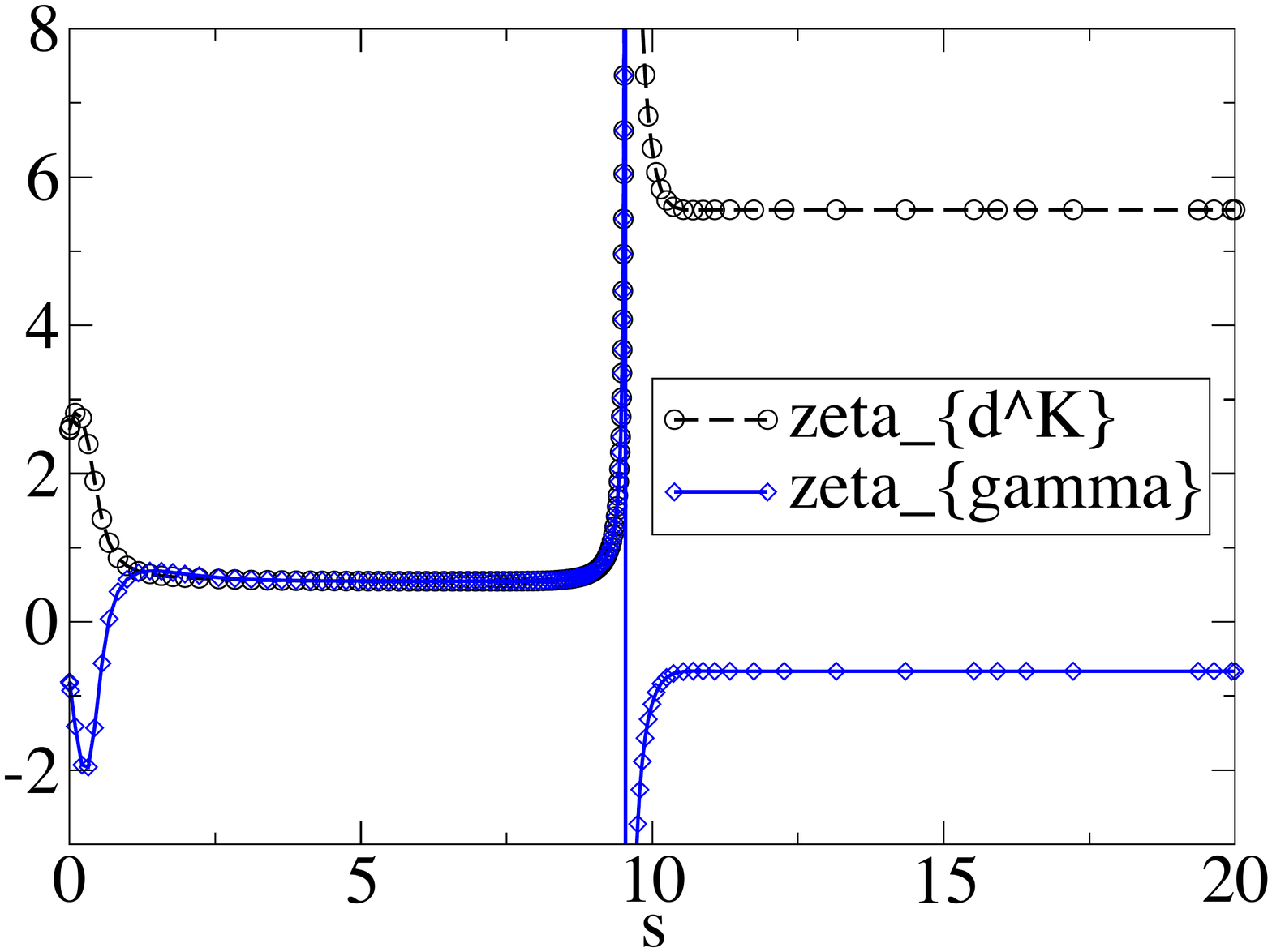}
\includegraphics[width=80mm]{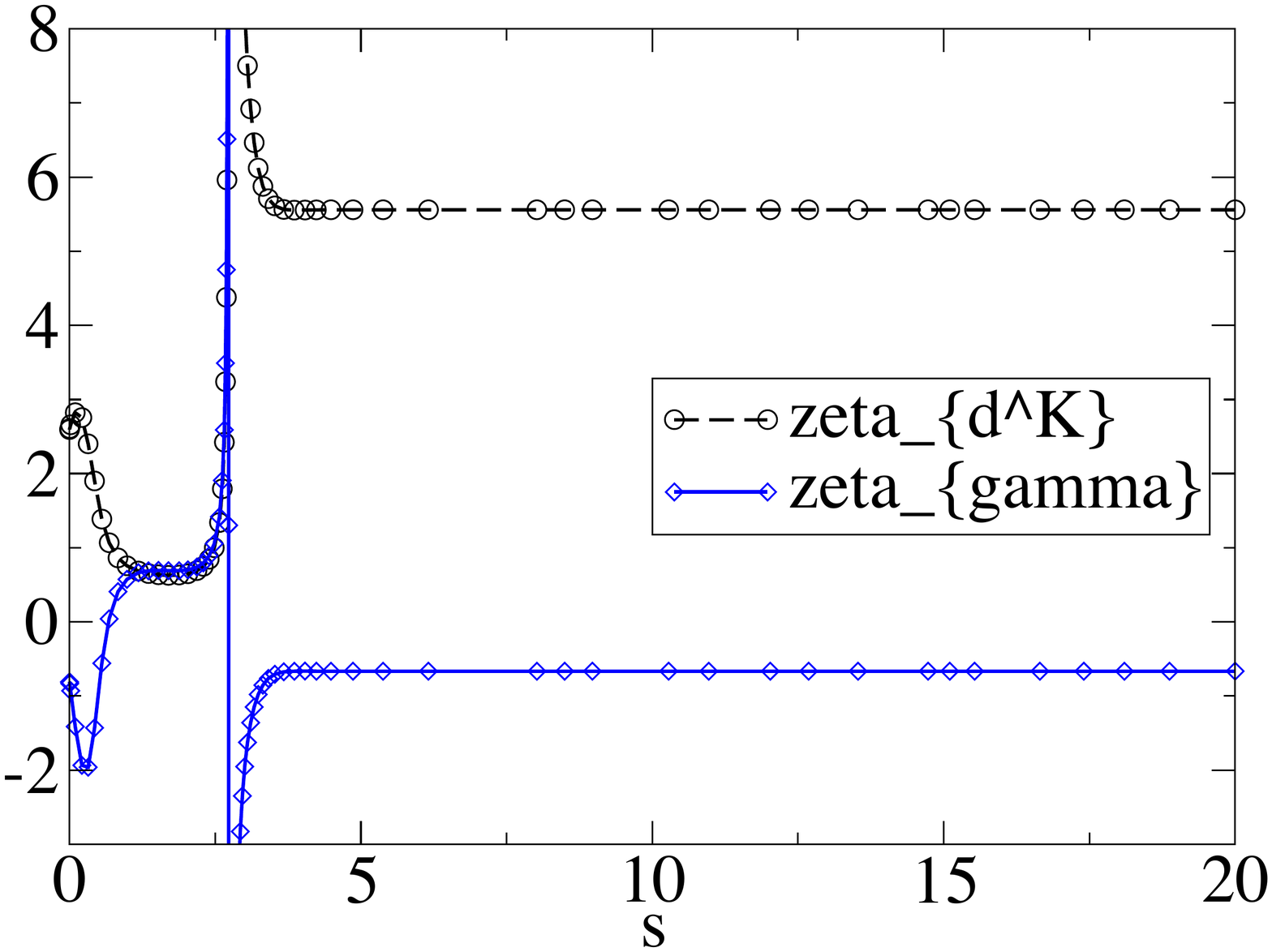}
\caption{Flow of the anomalous exponents slightly beyond the roughening transition 
in the rough phase in $d=3$. The unstable fixed point plateau in which both exponents 
take the same value becomes progressively shorter the deeper one goes into the rough phase. 
The flow goes 
from the UV (left of plot) to IR (right of plot) via $\Lambda= \Lambda_0 \exp(-s)$. 
Left: $\tilde{\lambda}_{\Lambda_0}= 6.44294680824$, 
Right: $\tilde{\lambda}_{\Lambda_0}= 6.6443$, both larger than $\tilde{\lambda}_{\Lambda_0,c}$. The scaling plateaus
of the roughening transition at which 
$\zeta_{d^K} = \zeta_{\gamma}=0.54$ for $d=3$ from Eq.~(\ref{eq:zetas_rt}) become unstable (at $s\approx 10$ in the left plot) 
and there is a steep transit into the rough phase, breaking the fluctuation-dissipation relation, 
at which $\zeta_{d^K}=5.56$ and $\zeta_{\gamma} = -0.66$ 
from Table \ref{tab:exponents_rough} in $d=3$. Fine-tuned to 13 digits the critical 
noise vertex for the roughening transition is $\tilde{\lambda}_{\Lambda_0,c}= 6.4429468082319$ with $\Lambda_0=10$, 
$\Delta_{\Lambda_0}=\tilde{\Delta}_{\Lambda_0}=0$ (no mass in the bare model Eq.~(\ref{eq:diffusion})).}
\label{fig:rtflows}
\end{figure}

The roughening transition and 
generically scale-invariant rough phase are distinguishable by different scaling forms of the response correlator:
\begin{align}
\mathcal{R}(\omega,\mathbf{k})&=-2 {\rm Im} \overline{\langle \phi(-\omega,-\mathbf{k}) \phi (\omega,\mathbf{k})\rangle}_{R} 
\quad\Rightarrow\quad
\mathcal{R}( s^z \omega, s \mathbf{k}) \propto \frac{1}{s^{2-\zeta_\gamma}} \mathcal{R}
\end{align}
and the independent Keldysh fluctuation correlator
\begin{align}
\mathcal{C}(\omega,\mathbf{k})& = i\, \overline{\langle \phi(-\omega,-\mathbf{k}) \phi (\omega,\mathbf{k})\rangle}_{K}
\quad\Rightarrow\quad
\mathcal{C}(s^z\omega, s \mathbf{k}) \propto \frac{1}{s^{4 - 2 \zeta_\gamma + \zeta_{d^K}}} \mathcal{C}\;,
\label{eq:keldysh_correlator}
\end{align}
where the overbar denotes the average over the random forces or noise. 
$\zeta_{\gamma}$ is the anomalous exponent for the linear time derivative in Eq.~(\ref{eq:diffusion}) and  
appears in the effective viscosity $\tilde{\nu} = \frac{\nu_0}{\gamma}$. $\zeta_{d^K}$ is the 
exponent for the effective noise spectrum, appearing in the statistical or Keldysh component defined below.
A finite value of the exponent $\zeta_{\rm hyper} = \zeta_{d^K} - \zeta_\gamma$ indicates deviations 
from thermal occupation of low-energy modes for which $\zeta_{\rm hyper} = 0$. 
This can be seen from the scaling form of the statistical distribution function
\begin{align}
f(\omega,\mathbf{k})= \frac{\mathcal{C}(\omega,\mathbf{k})}{\mathcal{R}(\omega,\mathbf{k})}
\quad\Rightarrow\quad
f(s^z\omega,s \mathbf{k}) \Rightarrow\frac{1}{s^{2+ (\zeta_{d^K} - \zeta_\gamma)}} \frac{\mathcal{C}}{\mathcal{R}}\;.
\end{align}
\begin{table}[t]
\centering
\begin{tabular}{c |  c c c}
\;\;\;\;   & \;\;\; \;\; $d=1$\;\;\;\;\;   & $\;\;\; \;\; $d=2$\;\;\;\;\;            $ & $\;\;\;\;\; $d=3$\;\;\;\;\;        $ \\ [0.5ex] 
\hline\\[-2mm]
$\chi_{\rm roughness}$ & $11.34$     & $5.15$    & $2.61$   \\[1ex]
\hline\\[-2mm]
$z_{\rm{dynamical}}$     & $8$           & $4$        & $2.66$   \\[1ex]
\hline\\[-2mm]
$\zeta_{d^K}$                & $15.68$    &  $8.30$   & $5.56$   \\[1ex]
\hline\\[-2mm]
$\zeta_\gamma$            & $-6$         & $-2$        & $-2/3$   \\[1ex]
\hline\\[-2mm]
$\zeta_{\rm hyper}$       & $21.68$    & $10.30$  & $6.23$   \\[1ex]
\hline\\[-2mm]
\end{tabular}
\caption{
One-loop values of critical exponents in the generically scale-invariant, rough phase. Explicit violation 
of a thermal fluctuation-dissipation relation is observed for which instead 
$\zeta_\gamma=\zeta_{d^K}$ and $\zeta_{\rm hyper}=0$.
The effective scale-dependent viscosity $\tilde{\nu}_\Lambda = 
\frac{\nu_{0}}{\gamma_\Lambda}\sim \Lambda^{\zeta_\gamma}$ diverges in entire the rough phase 
as $\zeta_\gamma < 0$. The fixed-point value of the nonlinear noise coupling goes 
to zero as $\epsilon = 4 - d  \rightarrow 0$ leading to vanishing $\zeta_{d^K}$, $\zeta_{\gamma}$ and $
z_{\rm dynamical}=2$ in $d=4$. Within our one-loop RG, $\epsilon$ may be regarded as 
the small parameter effectively controlling the flow; extrapolations to $d=1,2$ should 
be regarded as qualitative estimates only.}
\label{tab:exponents_rough}
\end{table}
These exponents are measurable in an interface experiment via the roughness exponent 
\begin{align}
\chi_{\rm roughness} = 1 - \frac{d}{2} + \frac{\zeta_{d^K} - \zeta_\gamma}{2}
\end{align}
which follows by comparison of Eq.~(\ref{eq:keldysh_correlator}) to the momentum representation 
of the height-height correlator 
$C_{hh}(\omega,\mathbf{q}) = \frac{1}{|\mathbf{q}|^{d+2 \chi +z}} \tilde{C} (\omega/|\mathbf{q}|^z)$ 
\cite{frey94} using further that in our case
\begin{align}
z_{\rm dynamical} = 2 - \zeta_\gamma\;.
\end{align}
For larger noise vertex $\tilde{\lambda}_{\Lambda_0}> \tilde{\lambda}_{\Lambda_0,c}$, 
in the rough or turbulent phase, the flow is attracted toward a gapless fixed point
which breaks the fluctuation-dissipation relation of the KPZ equation with white noise 
\cite{kpz86,frey94} as is shown in Fig.~\ref{fig:rtflows}. In particular, for $d=1$, the KPZ scaling relations 
$\chi + z = 2$ and $\chi=1/2$ are violated.
\begin{figure}[t]
\includegraphics[width=90mm]{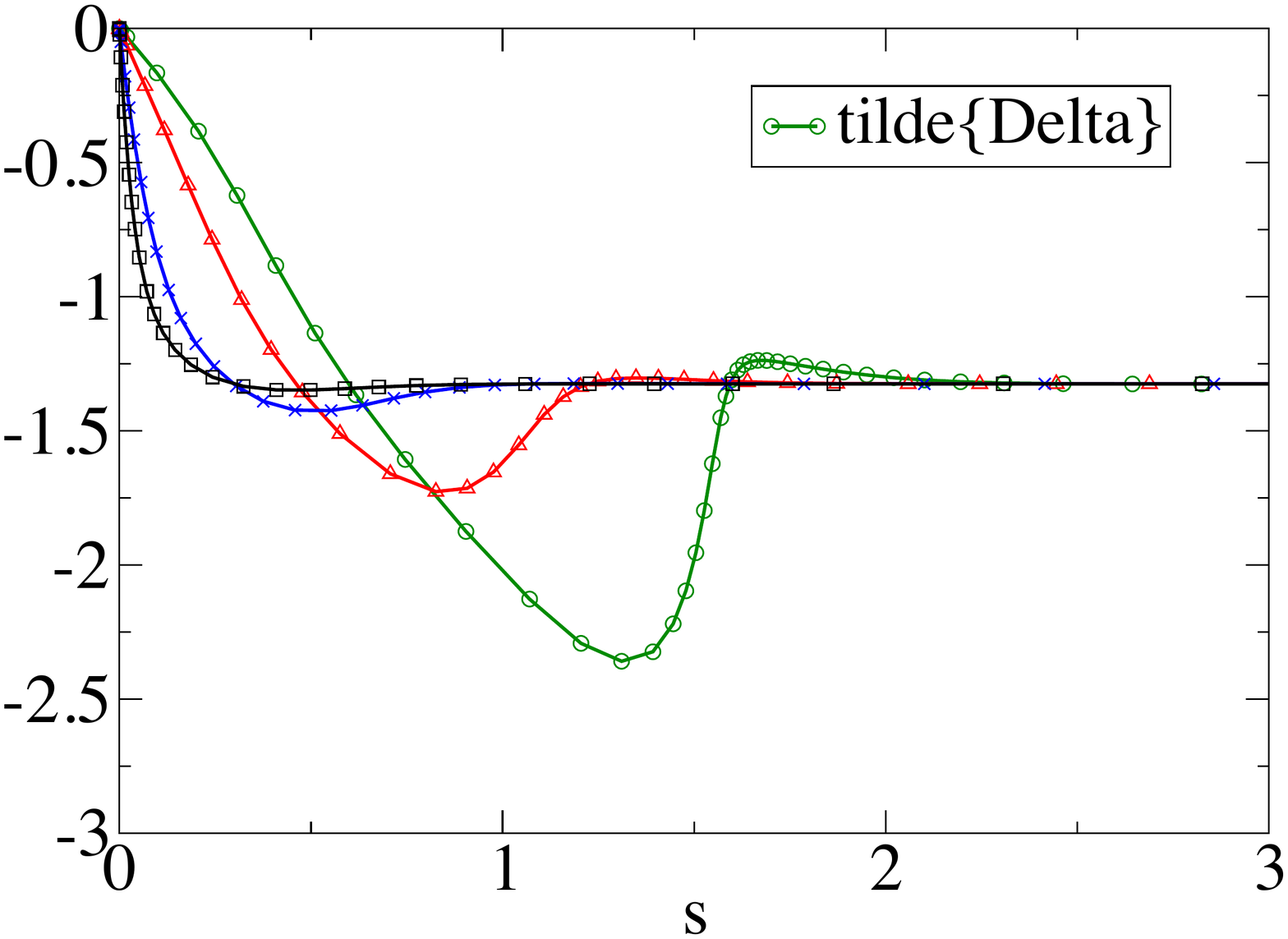}\\[-2mm]
\caption{Generically scale-invariant flows of the rescaled mass $\tilde{\Delta}_\Lambda$ in the rough phase 
in $d=3$, 
which is attracted to the {\it same} fixed point value
for different values of the noise vertex.
$\tilde{\lambda}_{\Lambda_0}=7,10,20,30$ from green circles (7) to 
black squares (30). The initially zero mass (no mass in bare model Eq.~(\ref{eq:diffusion})) is generated. The physical mass 
$\Delta_{\Lambda} = \tilde{\Delta}_{\Lambda} \Lambda^2 \gamma$ 
vanishes at the end of the flow for all couplings $\tilde{\lambda}_{\Lambda_0}\geq\tilde{\lambda}_{\Lambda_0,c}$. 
No fine-tuning of parameters required to reach this generically scale-invariant phase. 
The flow goes 
from the UV (left of plot) to IR (right of plot) via $\Lambda= \Lambda_0 \exp(-s)$. }
\label{fig:self-organized}
\end{figure}
The low-energy statistics 
in this phase is ``hyperthermal'', that is the low-energy mode power-law 
divergence is stronger than thermal with $\zeta_{\rm hyper}=6.23$ in $d=3$. Such infra-red enhanced 
population has been obtained at non-thermal fixed points of other, typically more complicated, 
field-theoretical models (see e.g. Refs.~\onlinecite{berges13,mathey14} and references therein).
The response function exponent turns out to be negative in the rough phase
\begin{align}
\zeta_{\gamma} = \frac{ 2(d-4)}{d}\;
\end{align}
and the full set of critical exponents in the rough phase are in Table~\ref{tab:exponents_rough}. 
The large value of the exponents in $d=1$, $d=2$ are due to the one-loop approximation 
and the relevance of the noise vertex for $d<4$. The one-loop computation 
is perturbatively controlled only close to $d=4$.

The rough/turbulent phase is generically scale-invariant (sometimes also referred to as self-organized critical) 
in the sense that it is does not require fine-tuning of the coupling constants to reach it beyond 
a certain threshold \cite{grinstein_review91,grinstein91}. Rather, 
the {\it same} fixed point is reached for all $\tilde{\lambda}_{\Lambda_0}> \tilde{\lambda}_{\Lambda_0,c}$
as can be seen from flows of the mass parameter in Fig.~\ref{fig:self-organized}. Note that the initial value of the mass $\Delta_{\Lambda_0}$ is always zero as 
appropriate for the gapless interface. Eq.~(\ref{eq:diffusion}) yields generic scale invariance with a relatively simple 
coupling to noise with a ubiquitous $1/f$-spectrum \cite{bak87}. 
A further appealing feature of the simple model Eq.~(\ref{eq:diffusion}) with $1/f$-noise 
is that the rough phase can be penetrated within a one-loop RG.

At the roughening transition (rt) in $d=2$ and $d=3$,
the asymptotically gapless dynamics fulfills an emergent thermal-like 
fluctuation-dissipation relation ($\zeta^{\rm rt}_{\rm hyper} = 0$) with
\begin{align}
\zeta^{\rm rt}_\gamma = \zeta^{\rm rt}_{d^K} = \frac{ 2 d }{ 8 + d}
\quad
\Rightarrow 
\quad
z^{\rm rt} = 2 - \zeta^{\rm rt}_\gamma = \frac{16}{8+d}
\label{eq:zetas_rt}
\end{align}
where $z^{\rm rt}$ is the dynamical exponent at one-loop. Consequently, the 
roughness exponent 
\begin{align}
\chi^{\rm rt}_{\rm roughness} = 1 - \frac{d}{2}
\end{align}
becomes negative for $d > 2$. In the massive phase, scaling stops completely and 
none of the above exponents are defined.
The roughening transition scaling is also seen in the unstable scaling plateaus 
of Fig.~\ref{fig:rtflows} and further numerical flows are presented in the main text, 
Subsec.~\ref{subsec:numerical}. At the roughening transition, 
the effective viscosity $\tilde{\nu}_{\Lambda} =\frac{\nu_0}{\gamma_{\Lambda}}\sim \Lambda^{\zeta_{\gamma}}$ 
vanishes as $\zeta^{\rm rt}_{\gamma} >0$ at one-loop. This is in contrast 
to the rough phase where the effective viscosity diverges as $\zeta_{\gamma} <0$ from Table \ref{tab:exponents_rough}.

\subsection{Organization of paper}
In Sec.~\ref{sec:BKPZ}, we recapitulate the relations between 
the Burgers, KPZ, and diffusion equation with multiplicative noise. 
Following Medina, Hwa, Kardar, and Zhang (MHKZ) \cite{medina89}, 
we show how temporal correlations in the noise break the Galilean invariance 
of the Burgers fluid. Then, we elevate the equation to an action on the closed-time 
Keldysh action in Subsec.~\ref{subsec:action}. In Subsec.~\ref{subsec:recap}, we briefly survey simplifications arising 
from the Galilean invariance such as exponent identities and a 
fluctuation-dissipation relation.

In Sec.~\ref{sec:RG}, we present the dynamic RG framework, explain the frequency cutoff technique 
and derive the form of the flow equations to one-loop order. 

In Sec.~\ref{sec:solutions}, we present analytical and numerical solutions to the flow equations.

In Sec.~\ref{sec:conclusions}, we offer some conclusions, point toward physical systems 
where our results may become relevant for, and outline potential directions for future work.

\section{Burgers-Kardar-Parisi-Zhang equation with $1/f$-noise}
\label{sec:BKPZ}
According to Kardar, Parisi, and Zhang \cite{kpz86}, coarse-grained fluctuations 
in the growth of a $d$-dimensional interface subject to random depositions 
can be described in terms a of scalar height function 
\begin{align}
\gamma \frac{\partial h}{\partial t} 
=
\nu_0 \nabla^2 h + \frac{\lambda}{2} \left(\nabla h\right)^2 + \eta\;,
\label{eq:kpz_original}
\end{align}
where both the height $h = h(t,\mathbf{x})$  and the noise $\eta = \eta(t,\mathbf{x})$ 
are functions of time $t$ and $d$-dimensional interface space spanned by $\mathbf{x}$.

Upon identifying the height with a vorticity-free velocity field $\mathbf{v}=-\nabla h$ and the 
random deposition noise with a random stirring force $\mathbf{f}= - \nabla \eta$,
Eq.~(\ref{eq:kpz_original}) is equivalent to the Burgers equation 
\begin{align}
\gamma \partial_t \mathbf{v} + \lambda \mathbf{v} \cdot \nabla \mathbf{v} = \nu_0 \nabla^2 \mathbf{v} + \mathbf{f}\;,
\label{eq:burgers}
\end{align}
where $\nu_0$ is the fluid viscosity, and the coefficient $\lambda$ parametrizes the relative strength of the 
nonlinear, convective term \cite{bouchaud95}.
In this paper, we will work with the representation of Eq.~(\ref{eq:kpz_original}) as 
a diffusion equation for $\phi(t,\mathbf{x})= \exp\left[\frac{\lambda}{2\nu_0 \gamma} h(t,\mathbf{x})\right]$ 
 \cite{kamenev_book} 
given above in Eq.~(\ref{eq:diffusion}). Under the Cole-Hopf transform
the scaling behavior of the correlators for the physical height variable $h$ and $\phi$ are proportional 
to each other
\begin{align}
C_{hh} = \langle h(\mathbf{x},t) h(\mathbf{x}',t') \rangle 
=
\frac{4 \nu_0^2 \gamma^2}{\lambda^2} 
\langle \ln \phi(\mathbf{x},t) \ln \phi(\mathbf{x}',t') \rangle
\sim
\frac{4 \nu_0^2 \gamma^2}{\lambda^2} C_{\phi\phi}\;.
\end{align}

\subsection{Broken Galilean invariance from $1/f$-noise}
The form of the noise correlator $\overline{\eta(t',\mathbf{x}') \eta(t,\mathbf{x})}$ 
in Eq.~(\ref{eq:pink})
determines the physical context and shapes the solution space 
of Eqs.~(\ref{eq:diffusion},\ref{eq:kpz_original},\ref{eq:burgers}). 
Temporal correlations in the noise break the 
Galilean invariance of the Burgers equation (\ref{eq:burgers}) under 
\begin{align}
\mathbf{v}(t,\mathbf{x}) \rightarrow \mathbf{v}_0 + \mathbf{v}'(\mathbf{x} - \lambda \mathbf{v}_0 t, t)
\label{eq:galilean}
\end{align}
associated with looking at the fluid in a moving frame \cite{forster77,medina89}. To be self-contained, we now recapitulate 
why this is so following App.~B of Ref.~\onlinecite{medina89}. For the interface equation (\ref{eq:kpz_original}) 
the Galilean invariance translates into invariance under infinitesimal tilts by a small angle $\mathbf{\epsilon}$
\begin{align}
h' &= h + \boldsymbol{\epsilon}\cdot \mathbf{x}
\nonumber\\
\mathbf{x}'&=\mathbf{x} + \lambda \boldsymbol{\epsilon} t'
\nonumber\\
t' &= t
\label{eq:tilt}
\end{align}
It is easy to see that the deterministic part of Eq.~(\ref{eq:kpz_original}) is invariant under Eq.~(\ref{eq:tilt}). It is also 
invariant under constant height shifts $h \rightarrow h + \rm{const}$ due to the absence of mass term or pinning 
potential. The transformed equation for $h'$ is subject to transformed noise 
$\eta'(t',\mathbf{x}') = \eta(t',\mathbf{x} + \lambda \boldsymbol{\epsilon} t')$ implying for the noise correlator
\begin{align}
F'
= \overline{\eta'(t'_1,\mathbf{x}'_1)\eta'(t'_2,\mathbf{x}'_2)}
&=
\overline{\eta(t_1,\mathbf{x}_1+\lambda \boldsymbol{\epsilon} t_1)\eta(t_2,\mathbf{x}_2+\lambda \boldsymbol{\epsilon} t_2)}
\nonumber\\
&=
F(t_1 - t_2, {\bf x}_1 - {\bf x}_2 + \lambda \boldsymbol{\epsilon} (t_1 - t_2))
\nonumber\\
&=
\delta(t_1 - t_2) F({\bf x}_1 - {\bf x}_2 + \lambda \boldsymbol{\epsilon}(t_1 - t_2))
=
\delta(t_1 - t_2) F({\bf x}_1 - {\bf x}_2) = F
\label{eq:noise_trafo}
\end{align}
where the last line is only true if the noise has no correlations in time 
i. e. $F(t,\mathbf{x}) = \delta(t) F(\mathbf{x})$. 
Our choice Eq.~(\ref{eq:pink}) corresponds to power-law correlations in real time and violates 
the invariance Eq.~(\ref{eq:noise_trafo}), as announced in the Introduction. 


\subsection{Keldysh path integral representation}
\label{subsec:action}

In order to explore the large distance and long time physics of the Burgers-Kardar-Parisi-Zhang 
systems subject to $1/f$-noise, we elevate the stochastic differential equation problem 
Eq.~(\ref{eq:diffusion}) plus Eq.~(\ref{eq:pink}) to a Keldysh path integral \cite{kamenev_book,dalladiehl13} 
on the closed time contour
\footnote{The main motivation for using Keldysh  
is to lay a basis for generalizing our work to quantum systems. 
There, one expects the short time, 
short distance behavior still be dominated by quantum effects and then 
the KPZ scaling to emerge in the IR at long times and large distances (see e.g.: 
Refs.~\onlinecite{kulkarni13,gangardt14}). To capture also the crossover scales, 
one needs to use Keldysh.}.
Wherever possible, we will follow the notation of Frey and T\"auber \cite{frey94}, 
who reviewed and performed this procedure and have given Ward 
and exponent identities for the related Janssen-De Dominicis functional.

The random forces are taken to be Gaussian-distributed
\begin{align}
W[\eta] \propto \exp
\left\{
- \int d^d x \int d \omega \frac{1}{2}
\eta(\omega,\mathbf{x})
|\omega|
\eta(\omega,\mathbf{x})
\right\}\;.
\label{eq:gaussian}
\end{align}
Then, the fluctuations in $\eta$ can be included on the same footing as the fluctuations 
of $\phi$ in the Keldysh generating functional
\begin{align}
\label{eq:noiseint}
Z  &= \int\mathcal D\eta W[\eta]\mathcal D (\phi,\tilde{\phi}) e^{i (S_\phi[\phi,\tilde{\phi}] - \int_{t,\mathbf{x}} \eta \phi\tilde{\phi})}
\nonumber\\
&\equiv 
\int\mathcal D(\eta,\phi,\tilde{\phi}) e^{i (S_\phi[\phi,\tilde{\phi}] + S_\eta[\eta]+S_{\lambda}[\phi,\tilde{\phi},\eta])}\;.
\end{align}
The momentum-independent noise propagator is now complex-valued
\begin{align}
S_\eta[\eta] &=  \frac{1}{2}\int_{\omega,\mathbf{q}}
\eta (-\omega,-\mathbf{q})\,\left[G^\eta(\omega)\right]^{-1} \,\eta (\omega,\mathbf{q}).
\label{eq:S_eta}
\end{align}
with 
\begin{align}\label{eq:geta}
G^\eta(\omega)=\frac{-i}{|\omega|}\;\;.
\end{align}
Note that the momentum integrations here is bounded in the UV by some short-distance cutoff by virtue 
of a necessary smallest physical distance below which the noise is spatially uncorrelated and 
the continuum description must be replaced by a discrete theory of lattice sites/grains.

The multiplicative noise term in Eq.~(\ref{eq:diffusion}) results in a trilinear noise 
vertex
\begin{align}
S_{\lambda}[\phi,\tilde{\phi},\eta] = 
- \int_t d t \int d^d x \,
\frac{\lambda}{2\nu_0} \,
\eta(t,\mathbf{x}) 
\tilde{\phi}(t,\mathbf{x}) 
\phi(t,\mathbf{x})
\label{eq:vertex}
\end{align}

Finally, the dynamics, diffusion, and statistics (Keldysh $\tilde{\phi}\tilde{\phi}$-component) 
of the $\phi$-fields are comprised in a matrix propagator
\begin{align}
S_{\phi}[\phi,\tilde{\phi}]=\frac{1}{2} \int d\omega \int d^d k \left( \phi \, \tilde{\phi} \right)
\left(\begin{array}{cc}
0 & \left[G^{A}(\omega,\mathbf{k})\right]^{-1}\\ 
\left[G^{R}(\omega,\mathbf{k}))\right]^{-1}   & D^K
\end{array}\right)
\left(\begin{array}{c}
\phi\\ 
\tilde{\phi}\\ 
\end{array}\right)
\label{eq:phase_keldysh} 
\end{align}
with the bare retarded and advanced Greens function given by
\begin{align}
G^R(\omega,\mathbf{k}) &=
 \frac{1}{ i  \gamma \omega - \nu_0 \mathbf{k}^2}\;,
\nonumber\\
G^A(\omega,\mathbf{k}) &= 
\frac{1}{ -i \gamma \omega -\nu_0 \mathbf{k}^2}\;,
\label{eq:prop_RA}
\end{align}
The statistical Keldysh component $D^K$ contains the effective noise spectrum
and will later be determined by loop corrections. Note that in absence 
of the initially zero $D^K$, the \emph{bare} action Eq.~(\ref{eq:vertex}) 
consists only of powers of $\tilde{\phi} \phi$, which can also be traced to a kind of 
gauge transformation related to Galilean invariance \cite{janssen99}. The bare action still 
describes a gapless interface and is invariant under constant shifts of $\phi\rightarrow \phi + {\rm const}$.
As we discussed, without Galilean invariance, the action is not protected against mass generation and 
we will allow for such terms to be generated in the renormalization group (RG) flow below in Sec.~\ref{sec:RG}.

\subsection{Mini-recap of known results without broken Galilean invariance}
\label{subsec:recap}

We here briefly recollect some previous results of the 
Burgers-KPZ field theories focussing in particular on 
the simplifications due to Galilean invariance in the case of temporally white noise.
In spatial dimension $d<2$ KPZ interfaces are always rough for any value of the 
nonlinearity $\lambda$, see e.g.: Ref.~\onlinecite{natter92}. For $d>2$, a smooth phase is stable 
for small $\lambda$ and there is a line of non-equilibrium roughening transitions 
separating the two. An important consequence of the ability to phrase the KPZ problem
in one dimension as an "equilibrium" partition function of an elastic string in a random potential 
\cite{huse85,kpz86} are fluctuation-dissipation relations \cite{deker75} and Ward identities \cite{frey94}. 
In particular, these insure that (i) the noise vertex $\lambda$ is not renormalized at all orders 
in perturbation theory, (ii) the fluctuation spectrum $D$ scales similarly to the dissipative 
viscosity $D/\nu_0 \rightarrow$ const, and (iii) that the roughness exponent $\chi$ and dynamical 
exponent $z$ fulfill the exact relation $\chi + z = 2$ with $z=3/2$ \cite{kpz86,medina89}.
In absence of Galilean invariance, (i) - (iii) do not hold anymore. In particular 
also the continuous shift invariance $h\rightarrow h + {\rm const}$ can now be violated by loop corrections.

To explore the interface 
dynamics constrained by only a reduced set conserved quantities (essentially only momentum and 
parity), we next perform a dynamic renormalization group analysis.

\newpage
\section{Dynamic renormalization group}
\label{sec:RG}
We will compute the one-loop RG flow of the action Eq.~(\ref{eq:noiseint}) 
employing a frequency cutoff, that is, rescaling frequencies and integrating over all 
momenta at each RG step. Our analysis will be framed in the context of the 
flow equation for the effective Keldysh action $\Gamma_{\Lambda}[\phi,\eta]$ 
as a function of a continuous flow 
parameter $\Lambda$ (see Refs.~\onlinecite{gezzi07,sieberer14} for 
condensed matter applications):
\begin{align}
\partial_{\Lambda}\Gamma_{\Lambda}[\phi,\eta] = \frac{i}{2} \text{Tr} \left[ \frac{ \dot{\mathcal{R}}}{ \Gamma_{\Lambda}^{(2)}[\phi,\eta] + \mathcal{R}} \right]
\;,
\label{eq:wetterich}
\end{align}
where the trace stands for a frequency and momentum integration and a simple matrix trace in field space over the 
$c$ and $q$ components, and the noise field $\eta$, respectively. 
$\mathcal{R}$ is a matrix containing cutoff functions (specified below)
as convenient for the field basis $ (\phi_c,\phi_q,\eta)$:
\begin{align}
\mathcal{R}=
\left(\begin{array}{ccc}
0 & R^{\phi}_{\Lambda} & 0 
\\ 
R^{\phi}_{\Lambda}&0&0
\\
0&0&R^{\eta}_{\Lambda}
\end{array}\right)
\nonumber
\end{align}
$\Gamma^{(2)}$ is a matrix containing the second field derivatives of $\Gamma$ 
evaluated at zero field whose inverse contains the scale-dependent Green's 
functions that we define below.

Before proceeding, let us mention previous works that employed 
the flow equation for the effective action, Eq.~(\ref{eq:wetterich}), 
for the KPZ problem with white \cite{canet10,kloss12, kloss14b} and spatially 
correlated noise \cite{kloss14}. These works highlighted the importance 
of the frequency and momentum dependence of the running couplings, especially 
to obtain quantitative estimates of critical exponents. We will see below to 
capture the qualitative flow for $1/f$ noise correlations, a one-loop truncation 
is sufficient. Once the difficulty of rescaling frequencies is overcome, the 
dynamical RG for the $1/f$ case is actually {\it simpler} than for the white noise case 
where no propagator renormalization occur to any order in the loop expansion 
\cite{kamenev_book}.

\subsection{Truncation of the effective action}
We now specify which parameters we keep out of the formally large set of 
coupling constants that can be generated under the RG flow. In particular, 
it will be important to introduce independent parameters for the response function and 
the Keldysh spectrum. That allows the flow to break equilibrium-like fluctuation-dissipation 
relations. We also introduce a mass term. Note that 
$\phi$, $\tilde{\phi}$ and $\eta$ all have bare scaling dimension $\left[ \frac{d}{2}\right]$ 
under bare $\omega \sim \Lambda^2$ power counting. 
As discussed in the Introduction, this makes the trilinear noise vertex $\lambda_\Lambda$ 
formally relevant in $d<4$. 

Propagator renormalizations are captured by introducing four flowing parameters 
$\gamma_\Lambda$, $A_\Lambda$, a mass term $\Delta_\Lambda$, and 
$d^{K}_\Lambda$ for the (independent) Keldysh component. Together with the noise 
vertex, this is also the minimal set of couplings that one would have to renormalize for 
example within a field-theoretic RG analysis \cite{frey94}. Including the additive, 
scale-dependent cutoff function $R_\Lambda$ into Eq.~(\ref{eq:phase_keldysh}) the 
quadratic part of the flowing action is 
\begin{align}
\Gamma^{(2)\phi}_{\Lambda}&=
\frac{1}{2} \int_{k} \left( \phi \, \tilde{\phi} \right)
\left(\begin{array}{cc}
0 & \left[G_\Lambda^{A}(\omega,\mathbf{k})\right]^{-1}\\ 
\left[G_\Lambda^{R}(\omega,\mathbf{k})\right]^{-1}   & D_\Lambda^K(\omega)
\end{array}\right)
\left(\begin{array}{c}
\phi\\ 
\tilde{\phi}\\ 
\end{array}\right)
\end{align}
with the now scale-dependent, retarded and advanced propagators
\begin{align}
G_\Lambda^R(\omega,\mathbf{k}) &= \frac{1}{ i \gamma_\Lambda \omega - \left(A_\Lambda \mathbf{k}^2 + \Delta_\Lambda\right) +
R^R_\Lambda(\omega) }\;,
\nonumber\\
G_\Lambda^A(\omega,\mathbf{k}) &= \frac{1}{ -i \gamma_\Lambda \omega - \left(A_\Lambda\mathbf{k}^2 +\Delta_\Lambda\right)
+ R^A_\Lambda(\omega) }\;,
\label{eq:prop_RA}
\end{align}
where the frequency cutoffs are complex conjugates of one another 
$R^A_\Lambda(\omega)=\left[R^R_\Lambda(\omega)\right]^\ast$ and defined below.
In contrast to the problem with full Galilean invariance, we will see that here a mass term 
(as well as nontrivial propagator renormalization via $\gamma_\Lambda$ discussed further below) 
are generated at one loop -- we capture this flow by introducing $\Delta_\Lambda$.

The Keldysh propagator $G^K = - G^R D^K G^A$ with
$D_\Lambda^K(\omega) = 2 i d_\Lambda^K$ is
\begin{align}
G_\Lambda^K(\omega,\mathbf{k}) 
= \frac{-2 i 
d^K_\Lambda}
{\left | i \gamma_\Lambda \omega - \left(A_\Lambda \mathbf{k}^2 + \Delta_\Lambda\right) +
R^R_\Lambda(\omega) \right |^2}\;.
\label{eq:prop_K}
\end{align}
At one-loop,  the momentum coefficient does not flow and it remains fixed at its initial value $A_{\Lambda} = \nu_0$ from 
Eq.~(\ref{eq:diffusion}).

\begin{figure} [t]
\includegraphics[width=150mm]{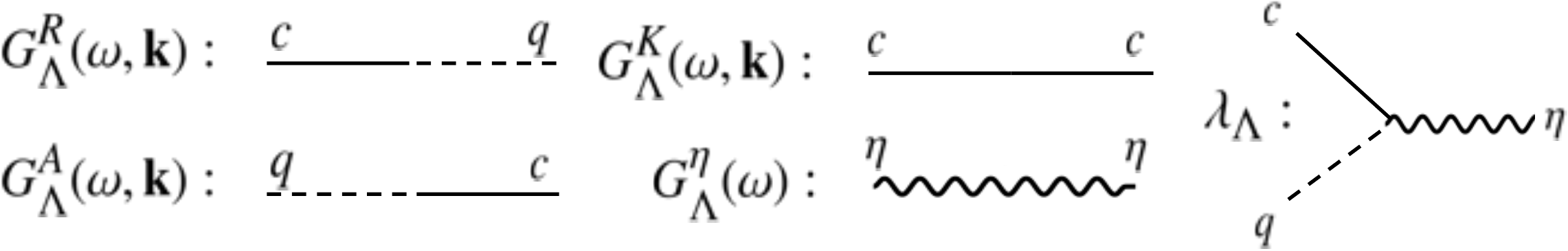}
\caption{Propagators and vertices appearing in the Keldysh action.}
\label{fig:diags}
\end{figure}

The flowing trilinear noise vertex
\begin{align}
\Gamma^{(3)}_{\Lambda}
&=
-\int_{t,\mathbf{x}}  
 \lambda_\Lambda 
\eta \tilde{\phi} \phi 
\end{align}
is related to the bare vertex Eq.~(\ref{eq:vertex}) at the beginning of the flow 
via $\lambda_{\Lambda=\Lambda_0} = \frac{\lambda}{2\nu_0}$.

The quadratic noise part in the action
\begin{align}
\Gamma_{\Lambda}^{(2)\eta}
&=
 \frac{1}{2}\int_{k}
\eta\,\left[G_\Lambda^\eta(\omega)\right]^{-1}\eta
\end{align}
 is not renormalized and the inverse propagator needs only be 
supplemented by the cutoff (also defined below)
\begin{align}
G^{\eta}_\Lambda(\omega)
=
\frac{-i}{|\omega| + R^\eta_\Lambda(\omega)}\;.
\label{eq:prop_eta}
\end{align}
By endowing the noise propagator with a cutoff, the noise average is performed 
continuously along $\Lambda$, which may be viewed as flowing 
from the short time dynamics at large $\Lambda$ to the long time 
dynamics at $\Lambda\rightarrow 0$. The Feynman graph elements 
of the flowing action are shown in Fig.~\ref{fig:diags}.

\subsection{Frequency cutoff technique}
As announced above, we will use frequency regulators with Eq.~(\ref{eq:wetterich}), 
which in Wilsonian RG language corresponds to rescaling frequencies
and integrating over all momenta at each RG step.

The specific form of the frequency regulator for the noise field is 
\begin{align}
R^\eta_\Lambda(\omega)& = 
\left(-|\omega| + \Lambda^2\right)
\theta\left[\Lambda^2 - |\omega|\right]
\nonumber\\
\partial_\Lambda R^\eta_\Lambda(\omega) 
&= 2 \Lambda \theta\left( \Lambda^2 - |\omega|\right)\;,
\end{align}
and for the $\phi$-field we have
\begin{align}
R^R_{\Lambda}(\omega) &= \gamma\left(-i \omega + i \rm{sgn}(\omega) \Lambda^2\right) 
\theta\left[\Lambda^2 - |\omega|\right]
\nonumber\\
\dot{R}^R_{\Lambda}(\omega)&\equiv\partial_\Lambda R^R_{\Lambda}(\omega) 
= 2\Lambda i \gamma  \rm{sgn}(\omega) \theta\left[ \Lambda^2 - |\omega|\right]
\end{align}
and for its advanced complex conjugate
\begin{align}
R^A_{\Lambda}(\omega) &= \gamma\left(+i \omega - i  \rm{sgn}(\omega) \Lambda^2\right) 
\theta\left[\Lambda^2 - |\omega|\right]
\nonumber\\
\dot{R}^A_{\Lambda}(\omega)&\equiv \partial_\Lambda R^A_{\Lambda}(\omega) =
-2\Lambda i \gamma  \rm{sgn}(\omega) \theta\left[\Lambda^2 - |\omega|\right]\;.
\end{align}
We also dropped, as usual, the higher-order scale derivatives  
$\partial_{\Lambda} \gamma$ in these expressions. Hard (e.g.\cite{strack08}) and soft (e.g. \cite{giering12}) 
frequency cutoffs are frequently also being applied in RG studies of 
strongly correlated fermionic systems.

It is well known that frequency regulators breaks the analyticity of propagators in the 
complex plane. In particular, unphysical contributions could be generated from loop 
integrations that would/should actually vanish: integrals with all poles in one half-plane, 
for example $\int d \omega \left( G^{R/A}(\omega) \right)^n$, which are identically zero 
upon closing the contour in the ``other'' half plane, would give a finite contribution with a 
frequency cutoff. In our flow equations below, such graphs do not appear at the 
one-loop level. Instead, we only encounter products of the noise propagator with 
the retarded and advanced propagators $\sim \int d \omega G^\eta(\omega) G^{R/A}(\omega)$, 
which give finite, and similar, contributions with or without frequency cutoffs
\footnote{Even if ``analytic'' contributions do occur in the flow, one just has to 
carefully remove them by hand, checking first if the loop integrations are finite 
without any cutoff.}.

Note that, as we discuss below, we also get around performing the frequency derivative of the cutoff 
to extract the flow of $\gamma$ by using an exponent identity that 
equates the flow of $\gamma$ with that of the noise vertex $\lambda$. 

With this cutoff choice, the frequency integrations over the one-loop contractions 
(below in Eq.~(\ref{eq:floweqs})) become simple 
due the cutoff choice and for the rotationally symmetric momentum integrations it is convenient 
to use a rescaled momentum variable $\tilde{\mathbf{k}}^2= \frac{A}{\gamma} \frac{\mathbf{k}^2}{\Lambda^2}$ 
such that the momentum integration measure becomes
\begin{align}
d ^d \mathbf{k} \rightarrow d^d{\tilde{\mathbf{k}}}\;\Lambda^d \left(\frac{\gamma}{A}\right)^{d/2}\;.
\end{align}
As announced above, we expect the continuum description to be valid only up to a short-distance 
of UV-momentum cutoff, below which the the granular/lattice structure of the interface becomes 
important. In order to regulate physically unimportant UV-divergences in the momentum 
integrations we systematically include only "on-shell" and smaller momenta into the flow
\begin{align}
\mathbf{k}^2 \leq \frac{\gamma}{A} \Lambda^2\;,
\quad
\tilde{\mathbf{k}}^2 \leq 1\;.
\end{align}
As $\Lambda\rightarrow 0$, this is a shrinking ball around the origin in momentum space, whose 
volume is continuously adapted as $\gamma$ flows, too. This mimicks the UV behavior from a 
Litim-type cutoff in momentum space, which regulates both, IR and 
UV divergences \cite{metzner_review12}. We have checked that the findings and fixed points 
reported below do not qualitatively seem to depend on the regularization procedure and 
choice of cutoffs. We have repeated the calculation for a different cutoff and different UV regularization 
procedure and found similar results
\footnote{
For example, we tried momentum cutoffs of the form 
$R^\phi_\Lambda(\mathbf{k}) = - A(\Lambda^2 - \mathbf{k}^2) \theta(\Lambda^2 - \mathbf{k}^2)$ 
and 
for the noise propagator $R^\eta_\Lambda = \Lambda^2$ as well as $R^\eta_\Lambda(\mathbf{k}) = 
\Lambda^2 \theta(\Lambda^2 - \mathbf{k}^2)$. The last choice ``entangles'' 
frequency of the noise with momenta of the $\phi$-field, but the flow was 
qualitatively similar to the ones with the frequency cutoff. We believe the 
using a frequency cutoff is cleaner and the ability to do so a particular strength of the 
1-PI functional flow equation (\ref{eq:wetterich}).
}

\subsection{One-loop flow equations}

We now write down the explicit form of the flow equations following from 
expanding the master flow equation, Eq.~(\ref{eq:wetterich}),  with the truncation specified above.
For brevity, we will use an integration symbol that includes the cutoff derivatives:
\begin{align}
\int
= \int \frac{d^d \mathbf{k}}{(2\pi)^d}
 \int \frac{ d\omega}{2\pi} \left[\dot{R}_{\eta}\partial_{R_\eta} 
 + \dot{R}^R_{\Lambda} \partial_{R^R_{\Lambda}}
 + \dot{R}^A_{\Lambda} \partial_{R^A_{\Lambda}}\;.
 \right]\;.
\end{align}
We get the expressions:
\begin{align}
%
%
 \partial_{\Lambda} (i d^K_\Lambda) &=
 \frac{-i}{2} 
 \int 
 \lambda_{\Lambda}^2
  G^\eta_\Lambda(\omega) 
G^K_\Lambda(\omega,\mathbf{k}) 
 \nonumber\\
\partial_{\Lambda} (-\Delta_\Lambda) &=
\frac{-i}{2} 
 \int 
\lambda_{\Lambda}^2 G^\eta_\Lambda(\omega)
   \left(G^A_\Lambda(\omega,\mathbf{k}) +G^R_\Lambda(\omega,\mathbf{k})\right)
 \nonumber\\
\partial_{\Lambda} (-\lambda_\Lambda) &=
\frac{-i}{2}\int  
\lambda_{\Lambda}^3
   G^\eta_\Lambda(\omega)
    \left(G^A_\Lambda(\omega,\mathbf{k}) ^2+G^R_\Lambda(\omega,\mathbf{k})^2\right)
%
%
   %
   %
\;.
\label{eq:floweqs}
\end{align}
\begin{figure} [t]
\includegraphics[width=140mm]{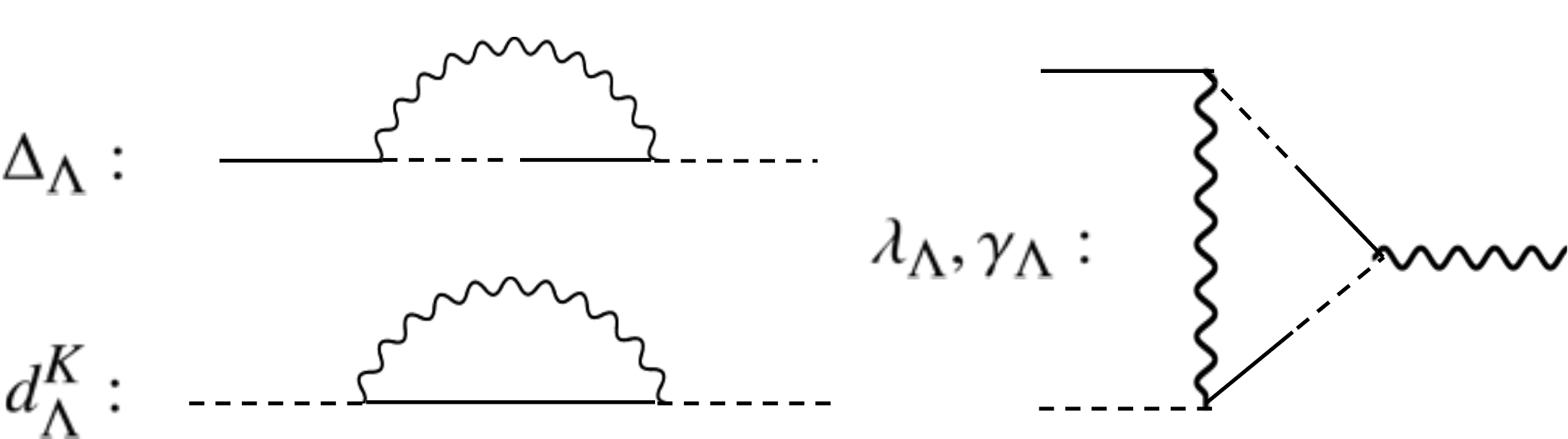}
\caption{One-loop contractions for Eq.~(\ref{eq:floweqs}). As for the flow of 
$\gamma_\Lambda$, the frequency derivative 
of the self-energy diagram for the retarded/advanced component (top left), 
generates the identical contraction as that for the noise vertex (right). This cancellation 
seems to be one reason for the generically scale-invariant nature of the rough phase as alluded 
to in the Introduction and observed in the numerics in Fig.~\ref{fig:self-organized}.}
\label{fig:contractions}
\end{figure}
The corresponding Feynman contractions are shown in Fig.~\ref{fig:contractions}.
At the one-loop level, there is no flow for the momentum renormalization factor $\partial_\Lambda A = 0$ and 
the flow of the frequency renormalization factor $\gamma_{\Lambda}$ is obtained via the diagrammatic identity
\begin{align}
\zeta_\gamma = 
-\frac{\Lambda}{\gamma_\Lambda} \partial_\Lambda \gamma_{\Lambda}
= 
-\frac{\Lambda}{\lambda_\Lambda} \partial_\Lambda \lambda_{\Lambda}
=
\zeta_\lambda\;
\label{eq:zetas}
\end{align}
also avoiding the necessity the perform frequency derivatives on the cutoff.

Severe diagrammatic redundancies appear in gauge theories (for example QED or 
the $CP^{N}$-model), there as a consequence of truly conserved global charges \cite{polyakov_book,huh2013}. 
Here, the cancellation is probably a leftover effect of the expansion of the action in powers of $\tilde{\phi} \phi$, which can be 
traced back to a gauge transformation related to the Galilean invariance for temporally white noise \cite{janssen99}. 

In addition to the anomalous exponents in Eq.~(\ref{eq:zetas}), we define 
the rescaled variables for the mass variable and the noise vertex
\begin{align}
\tilde{\Delta}&=\frac{\Delta_{\Lambda}}{\gamma_\Lambda \Lambda^2}
\nonumber\\
\tilde{\lambda}&=\frac{ \lambda_{\Lambda}}{\Lambda^{(4-d)/2} A^{d/4}_\Lambda 
\gamma^{1-d/4}_\Lambda\sqrt{\pi}}
\end{align}
and the slope of the statistical Keldysh component
\begin{align}
\zeta_{d^K} = - \frac{\Lambda}{d^K_\Lambda} \partial_\Lambda d^K_{\Lambda}\;.
\end{align}
For completeness, we will write out the analogous exponent 
for the momentum factor $\zeta_A = - \frac{\Lambda}{A_\Lambda} \partial_\Lambda A_{\Lambda}$ 
in the equations below, but it vanishes at the one-loop level.
%
%
%
%

The various $\beta$-functions following from explicit evaluation of Eqs.~(\ref{eq:floweqs}) 
take on the simple form:
\begin{align}
\Lambda\partial_\Lambda \tilde{\Delta}=& \left(-2 + \zeta_\gamma\right) \tilde{\Delta} + \tilde{\lambda}^2 D_{\lambda^2}[\tilde{\Delta}] 
\label{eq:betas}\\[2mm]
\Lambda\partial_\Lambda \tilde{\lambda} =& \left(\frac{d-4}{2} + \frac{d}{4} \zeta_A
+ (1 - \frac{d}{4}) \zeta_\gamma 
- \zeta_\lambda\right) \tilde{\lambda}
%
%
%
\nonumber
\end{align}
together with the anomalous exponents
\begin{align}
\zeta_A &= 0
\nonumber\\
\zeta_{d^K}& = \tilde{\lambda}^2 S_{\lambda^2}[\tilde{\Delta}] 
\nonumber\\
\zeta_\gamma & = \tilde{\lambda}^2 G_{\lambda^3}[\tilde{\Delta}] = \zeta_\lambda\;.
\label{eq:etas}
\end{align}
Invoking $\zeta_A=0$ and the identity $\zeta_{\gamma}=\zeta_{\lambda}$ the flow equation for the 
noise vertex simplifies to
\begin{align}
\Lambda\partial_\Lambda \tilde{\lambda} =& \left(\frac{d-4}{2} 
- \frac{d}{4} \zeta_\gamma 
\right) \tilde{\lambda}\;.
\label{eq:noise_simplified}
\end{align}
Partial cancellations in $\beta$-functions can be a reason for the appearance of 
anomalous exponents that depend only on dimensionality \cite{hwa89} and sometimes critical phases 
rather than critical points.

The fixed point structure is determined by dimensionality $d$ and the properties 
of the three threshold functions $D_{\lambda^2}[\tilde{\Delta}]$, $S_{\lambda^2}[\tilde{\Delta}]$, 
and $G_{\lambda^3}[\tilde{\Delta}]$, which depend on the mass variable $\tilde{\Delta}$ 
and dimensionality $d=1,2,3$. 
The analytic expressions and various limiting cases 
for the threshold functions are given in Appendix \ref{app:threshold}.

\section{Solving the flow}
\label{sec:solutions}
In this section, we solve the flow equations (\ref{eq:betas},\ref{eq:etas}) first analytically 
in two limit cases complementary to the numerical solutions exhibited in the key results 
section \ref{subsec:results}. The initial value for the mass 
variable is zero $\Delta_{\Lambda_0}=0$, having in mind a gapless interface before 
turning on the coupling to the noise. We will vary strength of the noise vertex 
$\lambda_{\Lambda}$ to tune through the phase diagram shown 
Sec.~\ref{subsec:results} from the massive phase (small initial $\tilde{\lambda}_{\Lambda_0}$) 
to the rough, but gapless phase (large $\tilde{\lambda}_{\Lambda_0}$) via critical point to 
the rough phase at $\tilde{\lambda}_{\Lambda_0,c}$. 


%
\subsection{Hyperthermal fixed point in the rough phase}
\label{subsec:rough}

\begin{figure} [t]
\includegraphics[width=85mm]{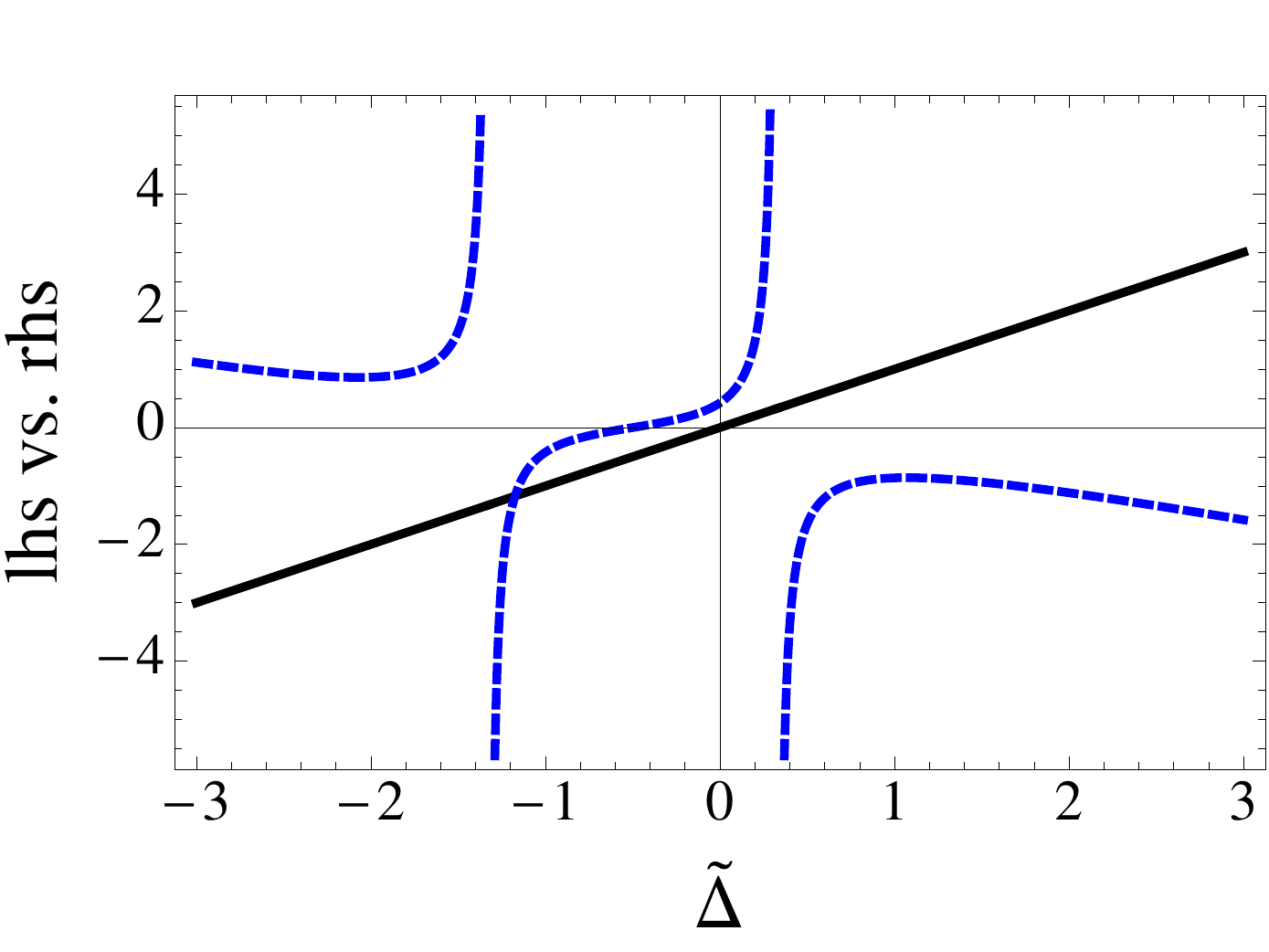}
\caption{Fixed point for $\tilde{\Delta}^\ast = -1.174$ in the rough phase in $d=2$
from finding intersections of the left-hand-side of 
Eq.~(\ref{eq:lhs_rhs}) (black line) with its right-hand-side (blue, dashed line). 
As at the Wilson-Fisher fixed point for the $O(n)$ model \cite{wilson71}, the rough phase fixed point, $(\tilde{\Delta}^\ast = -1.396, \tilde{\lambda}^\ast = 10.154)$ also in $d=3$, lies at negative mass in the plane of rescaled mass vs. rescaled coupling. This is true in all dimensions $d=1,2,3$.
}
\label{fig:fixed_rough}
\end{figure}

In $d<4$, Eqs.~(\ref{eq:etas},\ref{eq:noise_simplified}) admit a stable non-Gaussian fixed point ($\Lambda \partial_\Lambda \tilde{\lambda}=0$) solution
\begin{align}
\zeta_\gamma &= \frac{2(d-4)}{d}
\nonumber\\[2mm]
\tilde{\lambda}_\ast^2 &= \frac{2(d-4)}{d\,G_{\lambda^3}[\tilde{\Delta}_\ast]}
\end{align}
provided the equation for the mass ($\Lambda \partial_\Lambda \tilde{\Delta}=0$) has a solution
\begin{align}
\tilde{\Delta}_\ast &= \frac{d-4}{4} \frac{D_{\lambda^2}[\tilde{\Delta}_\ast]}{G_{\lambda^3}[\tilde{\Delta}_\ast]}\;
\label{eq:lhs_rhs}
\end{align}
such that $\tilde{\lambda}^2_\ast >0$ with $\tilde{\lambda}$ a real-valued number.
This is indeed the case in $d=1,2,3$ as is shown in Fig.~\ref{fig:fixed_rough} for $d=2$.
Of course, a fixed point $\tilde{\Delta}^\ast $ means that the physical mass 
vanishes during the flow
\begin{align}
\Delta = \tilde{\Delta}^\ast \Lambda^2 \gamma \underset{\Lambda\rightarrow 0} \rightarrow 0
\end{align}
implying that the entire rough phase is gapless. This fixed point is accompanied by a set of critical exponents including 
the dynamical exponent 
\begin{align}
z = 2 + \zeta_A - \zeta_\gamma = 2 - \zeta_\gamma\;
\end{align}
collected in Table \ref{tab:exponents_rough}.
In the rough phase, the strong-coupling between the "flat $z=\infty$ spectrum" (local in space) of the noise 
propagator and the bare $z=2$ overdamped $\phi$-dynamics leads to $z$-values intermediating between the two values.
Note that the steep flow when the 
scaling of the roughening transition changes to that of the rough phase in Fig.~\ref{fig:rtflows}, is due to 
a change of sign of the threshold function for $\zeta_{\gamma}$. This function,
$G_{\lambda^3}[\tilde{\Delta}]$, is plotted in the Appendix \ref{app:threshold}. Self-organized and generically scale-invariant phases in open systems have of course been discussed in a variety of contexts 
(see e.g. Hwa and Kardar \cite{hwa89} for a one-loop analysis of sandpile models and Ref.~\onlinecite{grinstein90} for a 
broader discussion on symmetries).

\subsection{Fixed point at the roughening transition}
\label{subsec:unstable}

\begin{figure} [b]
\includegraphics[width=100mm]{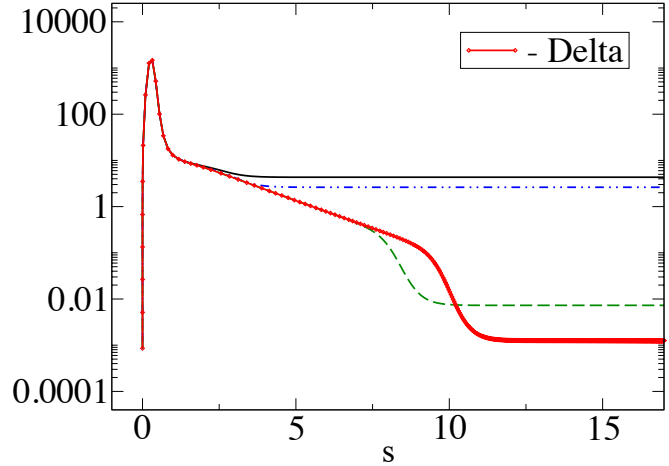}\\[-2mm]
\caption{Flow of the physical (non-rescaled) mass $-\Delta_\Lambda$ in $d=3$
upon approaching the roughening transition from the massive phase for 
different values of the initial noise vertex in 
double logarithmic graph (with $s = -\log[\Lambda/\Lambda_0]$ the flow goes 
from the UV (left of plot) to IR (right of plot)). 
The mass is initially numerically zero $\Delta_{s=0}=0$.
The red line is closest to the roughening transition $\tilde{\lambda}_{\Lambda_0,c}$ from below
while the black line has a $\tilde{\lambda}_{\Lambda_0}$ furthest away 
from $\tilde{\lambda}_{\Lambda_0,c}$. 
For infinite numerical accuracy the power-law linear scaling with slope $\zeta^{\rm rt}_{\gamma} = 6/11 = 0.54$ from 
Eq.~(\ref{eq:zetas_rt}) for the 
roughening transition would extend longer 
and longer.}
\label{fig:mass}
\end{figure}

The fixed point at the roughening transition (rt) is analyzed in changed 
variables
\begin{align}
\tilde{\Delta}_{\rm rt} &= \tilde{\Delta} \Lambda^2 \gamma^2
\nonumber\\
\tilde{\lambda}_{\rm rt} &= \tilde{\lambda}  \Lambda^2 \gamma^2
\label{eq:variable_change}
\end{align}
which attain fixed points at the roughening transition, ($\tilde{\Delta}^\ast_{\rm rt} = - 5.81$, $\tilde{\lambda}_{\rm rt}^\ast=23.33$) in $d=3$,
whose ratio turns out to be fixed to be $\left | \frac{\tilde{\lambda}^\ast_{\rm rt}}{\tilde{\Delta}^\ast_{\rm rt}}\right| = 3 \pi \sqrt{\frac{2}{11}}$. 
This leads to an automatic fulfillment of the vanishing 
of the $\beta$-function indicative of an asymptotically unstable fixed point. In the numerics, 
this is reflected by shorter and somewhat more wobbly scaling plateaus of the mass and vertex 
when compared to the stable fixed point of the rough phase. This means that the physical mass vanishes 
$\Delta_{\Lambda} = \tilde{\Delta}_{\rm rt}/\gamma \sim \Lambda^{\zeta^{\rm rt}_{\gamma}}$ as 
can be seen in Fig.~\ref{fig:mass}. 

For the suitably rescaled noise vertex, we can write the flow equation
\begin{align}
\partial_t \tilde{\lambda}_{\rm rt}
=
\left( \frac{d}{2} - (1 + \frac{d}{4}) \zeta_{\gamma_{\rm rt}} 
- \zeta_{\lambda_{\rm rt}}\right) \tilde{\lambda}_{\rm rt}\;,
\end{align}
which implies Eq.~(\ref{eq:zetas_rt}) (with $\zeta_{\gamma_{\rm rt}} = \zeta_{\lambda_{\rm rt}}$).
The emergent thermal fluctuation-dissipation relation is easily seen as follows. 
When $\tilde{\Delta}_{\rm rt}^\ast$ attains a constant fixed point value, 
the ``old'' variable $\tilde{\Delta}$ must diverge. Expanding the threshold functions
in Appendix \ref{app:threshold}
for the Keldysh component (in ``old'' variables and for concreteness in $d=3$)
\begin{align}
\zeta_{d^K}
=
\tilde{\lambda}^2 S_{\tilde{\lambda}^2}[\tilde{\Delta}]
\underset{\tilde{\Delta}\rightarrow \infty}{\rightarrow}
\approx
\tilde{\lambda}^2 \frac{1}{3 \pi^2 \tilde{\Delta}^2}
=
\tilde{\lambda}^2 G_{\lambda^3}[\tilde{\Delta}\rightarrow \infty]
= \zeta_{\gamma}\;.
\end{align}
Note that the ratio $\frac{\tilde{\lambda}^2}{\tilde{\Delta}^2}$ is invariant 
under the variable change Eq.~(\ref{eq:variable_change}) and 
consequently $\zeta_{\gamma}^{\rm rt} = \zeta_{d^K}^{\rm rt}$ as 
seen in the explicit flows in Subsec.~\ref{subsec:results} and Subsec.~\ref{subsec:numerical}.

\subsection{Numerical flows}
\label{subsec:numerical}
\begin{figure} [t]
\includegraphics[width=81mm]{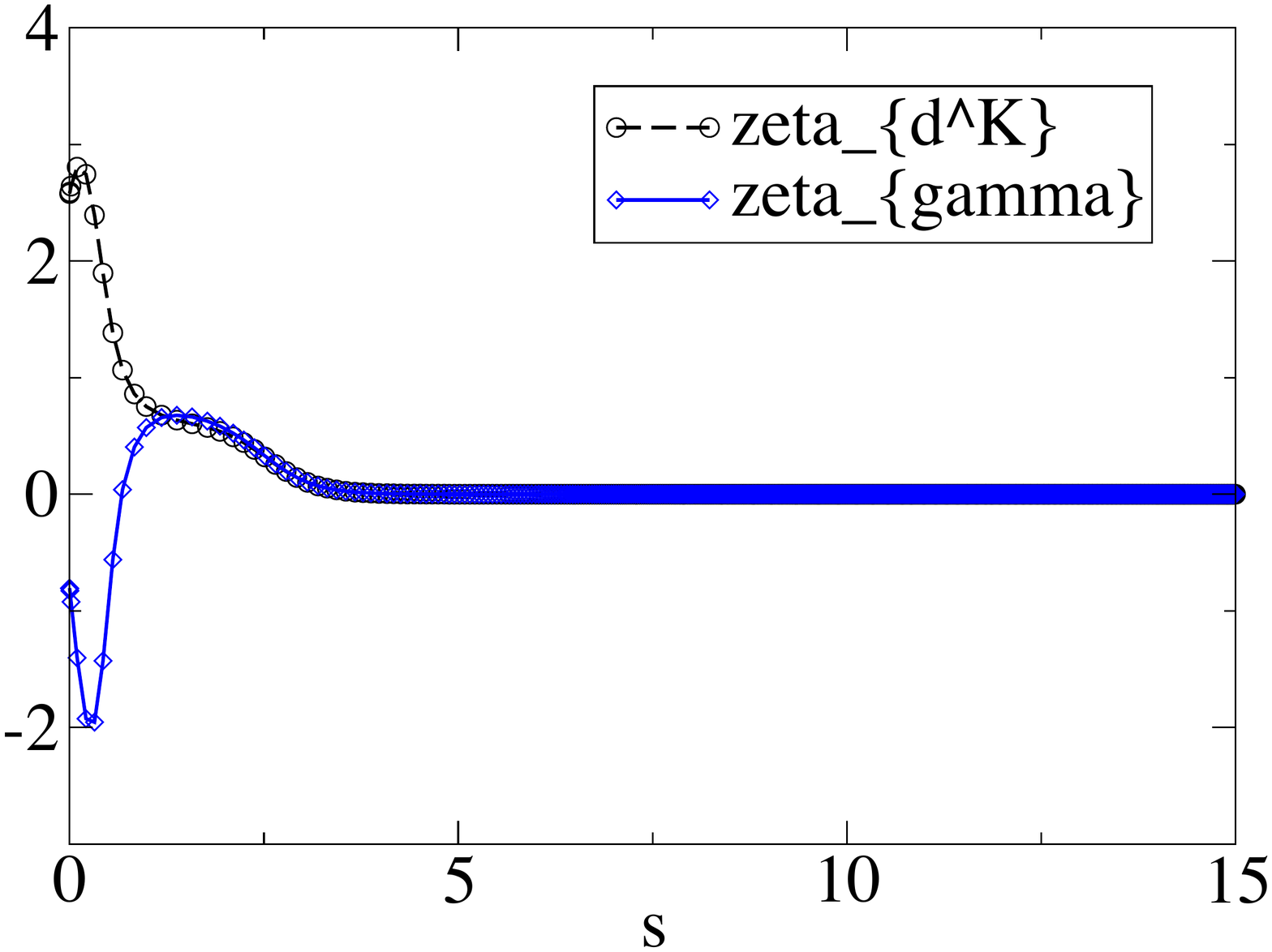}
\includegraphics[width=81mm]{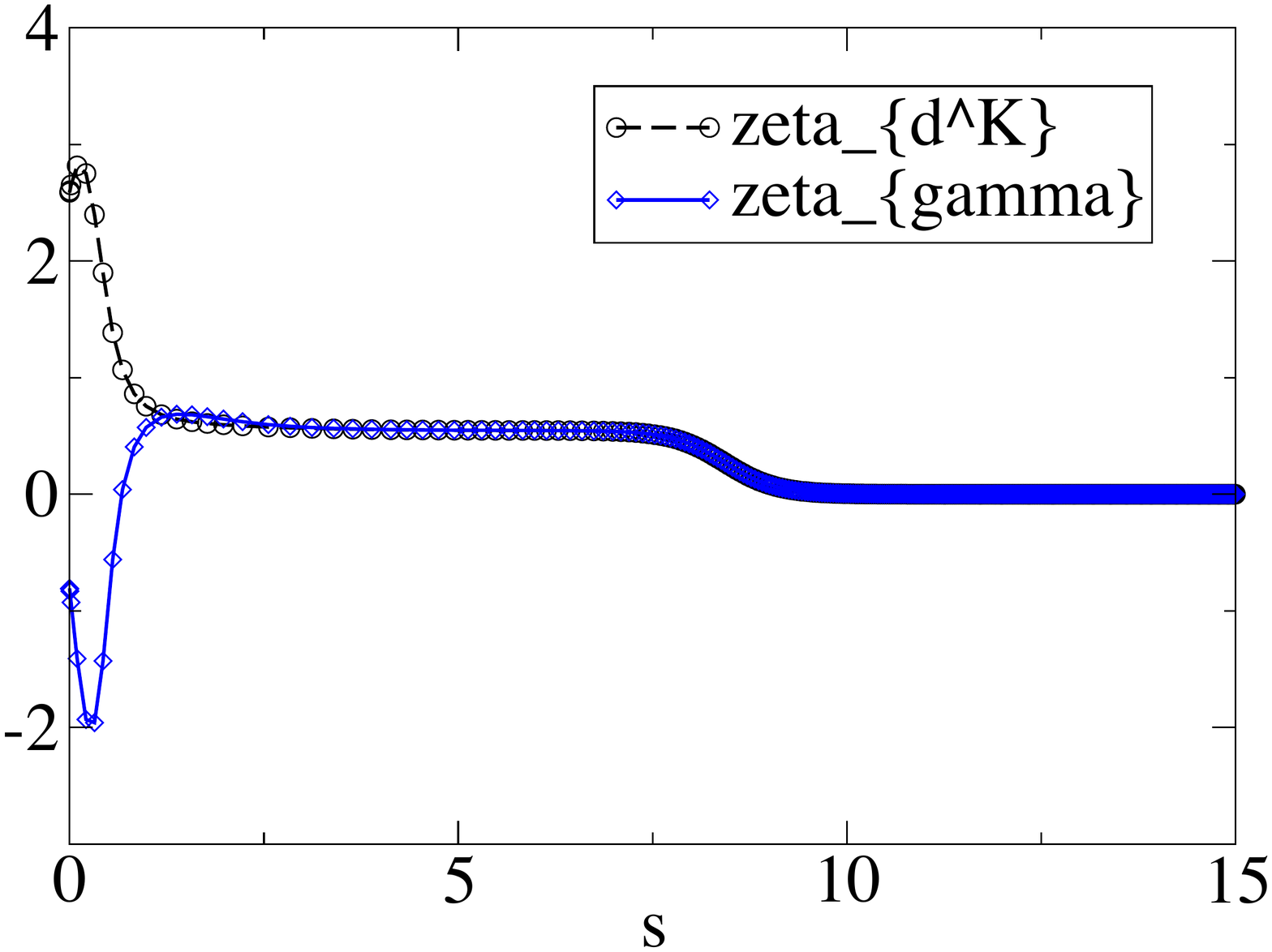}\\[-2mm]
\caption{Flow of the anomalous exponents when approaching the roughening transition 
from the massive phase. Left: $\tilde{\lambda}_{\Lambda_0}= 6.43$, 
Right: $\tilde{\lambda}_{\Lambda_0}= 6.442946808$. The closer the noise vertex 
is tuned to the critical value, the longer for the scaling plateau at which 
$\zeta_{d^K} = \zeta_{\gamma}=0.54$ for $d=3$ from Eq.~(\ref{eq:zetas_rt}). The flow goes 
from the UV (left of plot) to IR (right of plot) via $\Lambda= \Lambda_0 \exp(-s)$. At some point 
the scaling stops, both anomalous exponents become zero, because we are still in the massive phase.}
\label{fig:massive}
\end{figure}
\begin{figure}[h!]
\includegraphics[width=85mm]{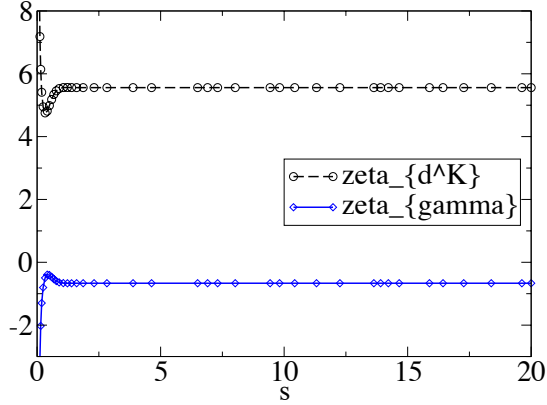}\\[-2mm]
\caption{Flows deep in the rough phase in which no 
remnant of the roughening transition (as in Fig.~\ref{fig:rtflows}) is visible and 
the anomalous exponents very quickly attain their rough phase fixed point values 
$\zeta_{d^K}=5.56$ and $\zeta_{\gamma} = -0.66$ 
from Table \ref{tab:exponents_rough} in $d=3$. Initial value of the noise vertex is 
$\tilde{\lambda}_{\Lambda_0}=30\gg \tilde{\lambda}_{\Lambda_0,c}= 6.4429468082319$. The flow goes 
from the UV (left of plot) to IR (right of plot) via $\Lambda= \Lambda_0 \exp(-s)$.}
\label{fig:deep_rough}
\end{figure}

We briefly describe the numerical procedure and initial conditions for the explicit 
flows in Subsec.~\ref{subsec:results}. We also show a few  
of more plots: Fig.~\ref{fig:mass},\ref{fig:massive} for flows in the massive phase upon approaching 
the roughening transition and Fig.~\ref{fig:deep_rough} as exemplary flow 
deep in the rough phase. The coupled flow equations (\ref{eq:betas},\ref{eq:etas}) are integrated using a fourth order Runge Kutta routine
(results did not change from using different routines) from 
high frequencies $\Lambda_0 = 10$ down to $\Lambda = 0$ using 
the momentum-integrated version of the threshold functions given in the Appendix \ref{app:threshold}. We always 
begin the flow with zero initial mass $\Delta_{\Lambda_0}=\tilde{\Delta}_0$ as appropriate for the gapless interface. 
The initial value of the rescaled noise vertex $\tilde{\lambda}_{\Lambda_0}$ is varied 
to obtain the phase diagram Fig.~\ref{fig:phasediag}. The condition for the phase boundary is 
the vanishing of the physical mass in the infrared $\Delta_{\Lambda\rightarrow0} \rightarrow 0$.

\section{Conclusions}
\label{sec:conclusions}

This paper pursued the strategy to (i) take an important universality class for 
non-equilibrium statistical mechanics (Burgers-Kardar-Parisi-Zhang equation), (ii) strip it
from an important conservation law(s) (Galilean invariance),  and (iii) 
compute the phase diagram and critical exponents using the dynamic 
renormalization group. 

We broke the Galilean invariance by 
accounting for temporal correlations in the random driving force.
The chief consequence of this is the absence of a fluctuation-dissipation relation 
even in $d=1$ for any noise level, and in $2\leq d<4$ for sufficiently strong noise levels. 
We penetrated this strong noise ``rough or turbulent'' phase within a dynamic RG 
flow using frequency rescaling techniques that took care of long-time correlations in the noise. 
We computed exponents to one-loop order, essentially controlled close to four 
space dimensions $d=4$. We showed that the rough phase is an example of generic scale 
invariance in the sense that its emergent gaplessness does not require fine tuning and traced 
this back to explicit cancellations in the one-loop $\beta$-functions. Higher loop analysis and 
numerical simulations will be needed to determine the fate of our theory beyond one-loop. 

Intriguing extensions of this work are quantum liquids and superfluids 
\cite{kulkarni13,gangardt14,altman13}
(in which more exotic types of noise can potentially be applied 
\cite{dalla12,buchhold14}), and waves in time-dependent random media 
\cite{segev12,saul92}. 
With regard to recent related works on 
nonlinear fluctuating hydrodynamics \cite{beijeren12,spohn13}, 
it will be interesting to systematically explore the role of broken
conservation laws onto crossover time-scales in 
dynamics and transport \cite{lux14}.

\acknowledgments
We are indebted to Sebastian Diehl for discussions and help with the renormalization group truncation
on the Keldysh contour. 
This work was supported by the DFG under grant Str 1176/1-1, by the Leibniz 
prize of A. Rosch, by the NSF under Grant DMR-1360789, by the Templeton foundation, 
by the Center for Ultracold Atoms (CUA), and by the Multidisciplinary University 
Research Initiative (MURI).

\appendix

\section{Threshold functions}
\label{app:threshold}

We here tabulate the threshold functions appearing in 
the flow equations (\ref{eq:betas},\ref{eq:etas}), 
$D_{\lambda^2}[\tilde{\Delta}]$, $S_{\lambda^2}[\tilde{\Delta}]$, 
and $G_{\lambda^3}[\tilde{\Delta}]$. We will give the pre-momentum 
integrated expression for general dimension $d$ and the post-momentum 
integrated expression only for $d=2$; the other dimensions do not qualitatively change 
their form.

\begin{figure} [b]
\includegraphics[width=75mm]{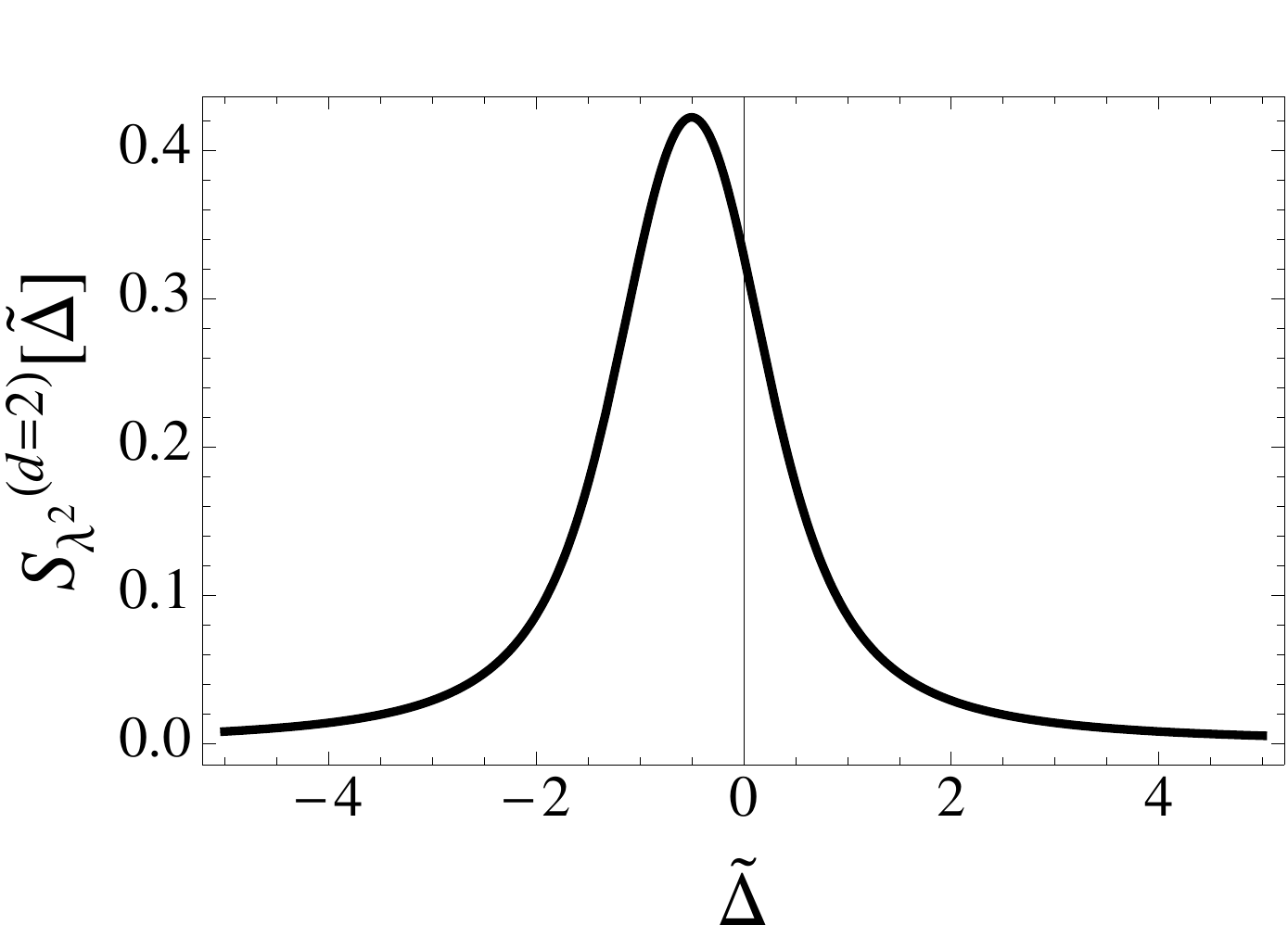}
\quad\quad
\includegraphics[width=79mm]{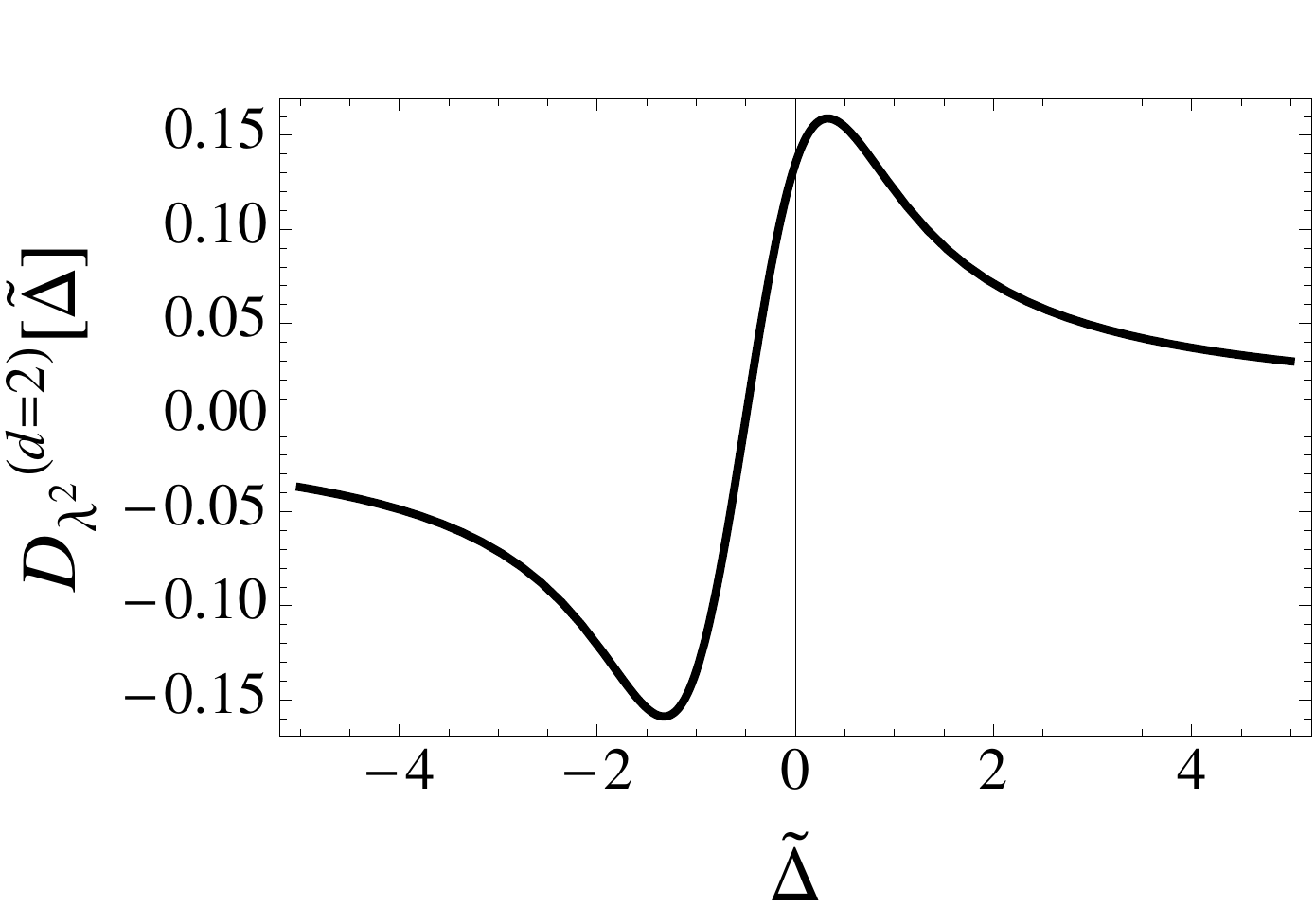}
\caption{Plots of the threshold functions for the Keldysh component (left) and the 
mass (right) in $d=2$.}
\label{fig:threshold_keldysh}
\end{figure}

The flow of the Keldysh component is determined by
\begin{align}
&S_{\lambda^2}^{(d)}[\tilde{\Delta}] = \int \frac{ d^d \tilde{k}}{(2\pi)^d}
\frac{2 \left(\left(\tilde{\Delta} +\tilde{\mathbf{k}}^2\right)^2+3\right)}
{\left(\left(\tilde{\Delta} +\tilde{\mathbf{k}}^2\right)^2+1\right)^2}
\Bigg |_{0\leq\tilde{\mathbf{k}}^2\leq 1}
\nonumber\\
&\overset{d=2}=
\frac{-2 \left(\tilde{\Delta} ^2+1\right) (\tilde{\Delta}  (\tilde{\Delta} +2)+2) \tan ^{-1}(\tilde{\Delta} )+2 \left(\tilde{\Delta}
   ^2+1\right) (\tilde{\Delta}  (\tilde{\Delta} +2)+2) \tan ^{-1}(\tilde{\Delta} +1)-\tilde{\Delta}  (\tilde{\Delta} +1)+1}{2 \pi 
   \left(\tilde{\Delta} ^2+1\right) (\tilde{\Delta}  (\tilde{\Delta} +2)+2)}\;.
\end{align}

The flow of the mass variable is determined by
\begin{align}
D_{\lambda^2}^{(d)}[\tilde{\Delta}]& = \int \frac{ d^d \tilde{k}}{(2\pi)^d}
\frac{2 \left(\tilde{\Delta} +\tilde{\mathbf{k}}^2\right) \left(\left(\tilde{\Delta}
   +\tilde{\mathbf{k}}^2\right)^2+3\right)}{\left(\left(\tilde{\Delta} +\tilde{\mathbf{k}}^2\right)^2+1\right)^2}
   \Bigg |_{0\leq\tilde{\mathbf{k}}^2\leq 1}
\nonumber\\
&\overset{d=2}=
\frac{\frac{4 \tilde{\Delta} +2}{\left(\tilde{\Delta} ^2+1\right) (\tilde{\Delta}  (\tilde{\Delta} +2)+2)}+\log
   \left(\frac{2 \tilde{\Delta} +1}{\tilde{\Delta} ^2+1}+1\right)}{4 \pi }\;.
\end{align}

The flow of the noise vertex (and the frequency renormalization factor) is determined by
\begin{align}
G_{\lambda^3}^{(d)}[\tilde{\Delta}]& = \int \frac{ d^d \tilde{k}}{(2\pi)^d}
\frac{2 \left(\left(\tilde{\Delta} +\tilde{\mathbf{k}}^2\right)^2 \left(\left(\tilde{\Delta}
   +\tilde{\mathbf{k}}^2\right)^2+6\right)-3\right)}{\left(\left(\tilde{\Delta} +\tilde{\mathbf{k}}^2\right)^2+1\right)^3}
      \Bigg |_{0\leq\tilde{\mathbf{k}}^2\leq 1}
   \nonumber\\
&\overset{d=2}=
\frac{\tilde{\Delta}  (\tilde{\Delta} +1) \left(\tilde{\Delta} ^2+\tilde{\Delta} +1\right) 
\left(\tilde{\Delta}^2+\tilde{\Delta}+6\right)-4}
{2 \pi\left(\tilde{\Delta} ^2+1\right)^2 (\tilde{\Delta}  (\tilde{\Delta} +2)+2)^2}\;.
\end{align}
\begin{figure} [t]
\includegraphics[width=80mm]{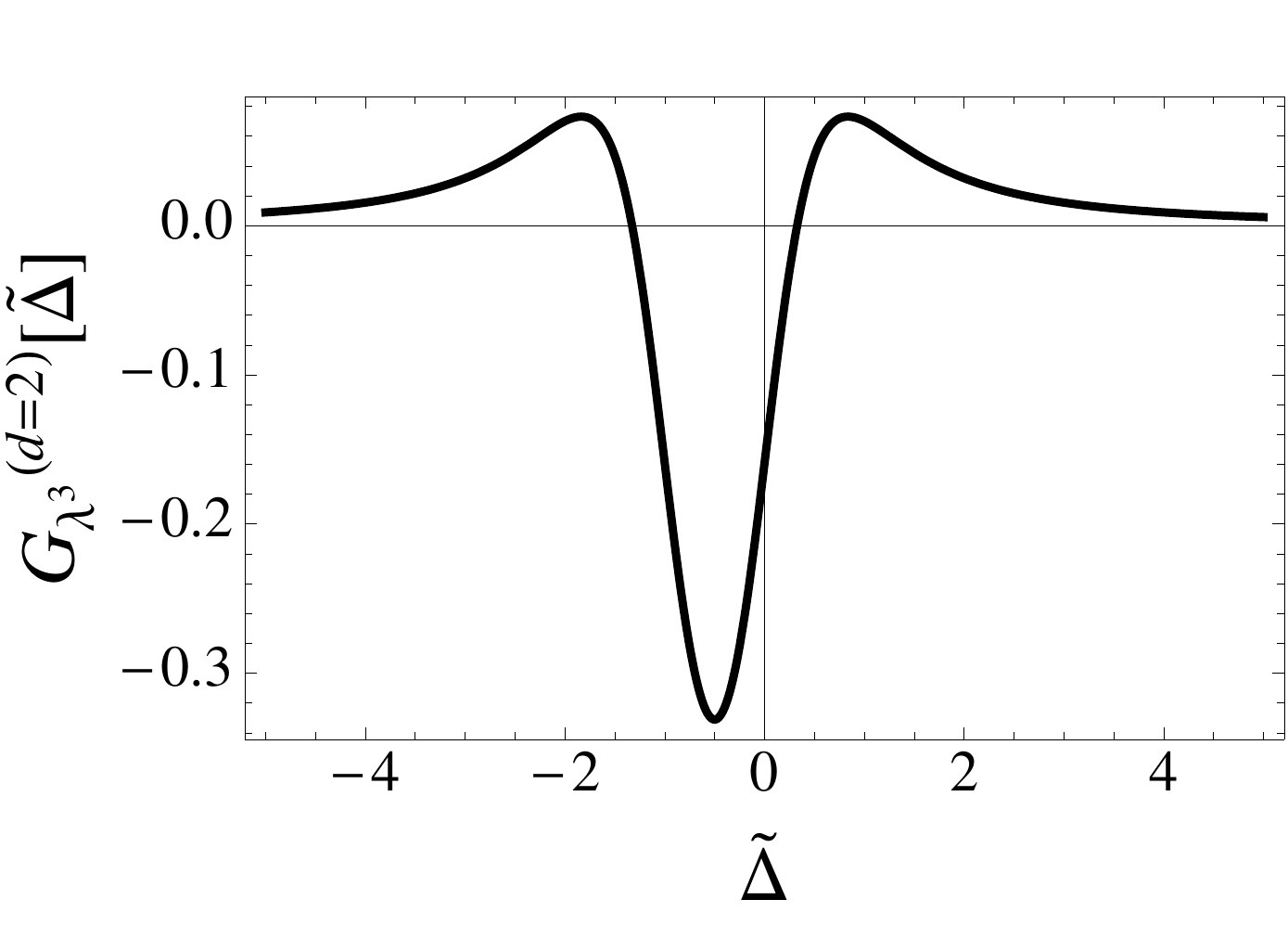}
\caption{Plot of the threshold functions for the noise vertex/frequency renormalization factor
in $d=2$.}
\label{fig:threshold_keldysh}
\end{figure}
%


\end{document}